\documentclass[12pt,preprint]{aastex61}
\usepackage{natbib}
\usepackage{hyperref} 
\usepackage{amsmath}
\newcommand{\um}{$\mu$m}

\newcommand{\Msun}{M$_{\odot}$}

\newcommand{\Lsun}{L$_{\odot}$}

\newcommand{\kms}{km~s$^{-1}$}

\newcommand{\hii}{\mbox{$\mathrm{H\,{\scriptstyle {II}}}$}}

\newcommand{\nthp}{N$_{2}$H$^{+}$\,(1--0)}

\newcommand{\hcop}{HCO$^{+}$\,(1--0)}

\newcommand{\sio}{\mbox{SiO~(2--1)}}

\newcommand{\hcnnt}{HCN}

\newcommand{\hnc}{HNC~(1--0)}
	
\newcommand{\hcn}{HCN (1--0)}

\newcommand{\cch}{C$_{2}$H\,(1--0)~3/2--1/2}

\newcommand{\cii}{[\mbox{$\mathrm{C\,{\scriptstyle {II}}}$}]}
\newcommand{\oi}{[\mbox{$\mathrm{O\,{\scriptstyle {I}}}$}]}
\newcommand{\lcii}{$L_{[C~II]}$}
\newcommand{\lfir}{$L_{FIR}$}

\newcommand{\sourceA}{AGAL302.496$-$0.031}
\newcommand{\sourceB}{AGAL313.576$+$0.324}
\newcommand{\sourceC}{AGAL317.429$-$0.561}
\newcommand{\sourceD}{AGAL341.702$+$0.051}

\shorttitle{Characterizing \cii\, Line Emission in Massive Star Forming Clumps}
\shortauthors{Jackson et al.}

\begin{document}
\received{6 April 2020}
\revised{6 August 2020}
\accepted{17 September 2020}
\submitjournal{ApJ}

\title{Characterizing \cii\, Line Emission in Massive Star Forming Clumps}

\author{James M. Jackson}
\affiliation{USRA SOFIA Science Center, NASA Ames Research Center, Moffett Field, CA 94045, USA}
\affiliation{School of Mathematical and Physical Sciences, University of Newcastle, University Drive, Callaghan NSW 2308, Australia}
\affiliation{Institute for Astrophysical Research, Boston University, Boston MA 02215, USA}

\author{David Allingham}
\affiliation{School of Mathematical and Physical Sciences, University of Newcastle, University Drive, Callaghan NSW 2308, Australia}

\author{Nicholas Killerby-Smith}
\affiliation{School of Mathematical and Physical Sciences, University of Newcastle, University Drive, Callaghan NSW 2308, Australia}
\affiliation{Research School of Astronomy and Astrophysics, Australian National University, Acton ACT 2601, Australia}

\author{J. Scott Whitaker}
\affiliation{Physics Department, Boston University, 590 Commonwealth Ave., Boston, MA 02215, USA}

\author{Howard A. Smith}
\affiliation{Harvard-Smithsonian Center for Astrophysics, 60 Garden St. Cambridge MA, USA}

\author{Yanett Contreras}
\affiliation{Leiden Observatory, Leiden University, PO Box 9513, 2300 RA Leiden, The Netherlands}

\author{Andr{\'e}s E. Guzm{\'a}n}
\affiliation{National Astronomical Observatory of Japan, National Institute of Natural Sciences, 2-21-1 Osawa, Mitaka, Tokyo 181-8588, Japan}

\author{Taylor Hogge}
\affiliation{Institute for Astrophysical Research, Boston University, Boston MA 02215, USA}

\author{Patricio Sanhueza}
\affiliation{National Astronomical Observatory of Japan, National Institute of Natural Sciences, 2-21-1 Osawa, Mitaka, Tokyo 181-8588, Japan}

\author{Ian W. Stephens} 
\affiliation{Harvard-Smithsonian Center for Astrophysics, 60 Garden St. Cambridge MA, USA}

\begin{abstract}
Because the 157.74 \um\, \cii\, line is the dominant coolant of star-forming regions, it is often used to infer the global star-formation rates of galaxies. By  characterizing the \cii\, and far-infrared emission from nearby Galactic star-forming molecular clumps, it is possible to determine whether extragalactic \cii\, emission arises from a large ensemble of such clumps, and whether \cii\, is indeed a robust indicator of global star formation.  We describe \cii\, and far-infrared observations using the FIFI-LS instrument on the SOFIA airborne observatory toward four dense, high-mass, Milky Way clumps.  Despite similar far-infrared luminosities, the \cii\, to far-infrared luminosity ratio, \lcii$/$\lfir, varies by a factor of at least 140 among these four clumps.  In particular, for \sourceB, no \cii\, line emission is detected despite a FIR luminosity of 24,000 \Lsun.  \sourceB\, lies  a factor of more than 100 below the empirical correlation curve between \lcii$/$\lfir\, and $S_\nu (63  \mu m)/S_\nu (158  \mu m)$ found for galaxies.  \sourceB\, may be in an early evolutionary ``protostellar" phase with insufficient ultraviolet flux to ionize carbon, or in a deeply embedded ``hypercompact' \hii\, region phase where dust attenuation of UV flux limits the region of ionized carbon to undetectably small volumes.  Alternatively, its apparent lack of \cii\, emission may arise from deep absorption of the \cii\, line against the 158 \um\, continuum, or self-absorption of brighter line emission by foregroud material, which might cancel or diminish any emission within the FIFI-LS instrument's broad spectral resolution element ($\Delta V \sim 250$ \kms).
 \end{abstract}


\section{Introduction}
The 157.74 $\mu$m $^{2}P_{3/2} - ^{2}P_{1/2}$ fine structure line of singly-ionized atomic carbon (hereafter ``\cii'') is one of the most luminous spectral lines in  gas-rich galaxies, typically containing ~0.0l\% to 1\% of a galaxy's total far-infrared (hereafter ``FIR'') luminosity (\lfir) (e.g., \citealt{Stacey2010}). Since young high-mass stars produce ultraviolet radiation that can ionize carbon, the \cii\,  line luminosity (\lcii) might be expected to be related to the star-formation rate. Indeed, \lcii\,  and the star-formation rate, as traced by FIR emission \citep[e.g.,][]{Luhman1998}, appear to be linearly correlated over 6 orders of magnitude, from the few pc scale of individual Galactic photodissociation regions (PDRs), to kpc scale regions in the Milky Way, and on to entire galaxies \citep{Pineda2014}. Because the \cii\,  line is so luminous, it can be detected  toward distant, high-redshift galaxies and potentially used as a probe of global star-formation in the early universe \citep[e.g.,][]{Kimball2015, Capak2015, Gullberg2015} as well as in more nearby galaxies \citep[e.g.,][]{ Stacey2010, Sargsyan2012}. 

However, the interpretation of the \cii\,  line and the FIR continuum emission from galaxies remains puzzling. For normal galaxies with \lfir\, $\sim 10^{10}$ \Lsun, the \cii\,  to FIR line-to-continuum luminosity ratio is \lcii$/$\lfir\, $\sim 3 \times 10^{-3}$ (and higher for lower metallicity galaxies, (e.g, see review by \citealt{Madden2000}).  However, at the higher FIR luminosities found in Luminous InfraRed Galaxies (LIRGs: \lfir\, $>10^{11}$ \Lsun) and Ultra-Luminous InfraRed Galaxies (ULIRGs: \lfir\, $ >10^{12}$ \Lsun), the \lcii$/$\lfir\, ratio steadily decreases as a function of \lfir\, by almost two orders of magnitude from that of normal galaxies, a result commonly known as the “\cii\,  deficit” \citep[e.g.,][]{ Ibar2015}.  Several explanations for the \cii\,  deficit have been posed, such as (1) the increased production of charged dust grains when the interstellar medium is exposed to high UV fields, which in turn inhibits the production of photoelectrons that collisionally excite ionized carbon, (2) an  increased rate of \cii\, collisional de-excitation at high densities, (3) unrelated IR contamination from evolved stars or AGN, and (4) absorption or self-absorption of the \cii\,  line (see \citealt{Ibar2015} and references therein). \cite{Ibar2015} favor the first of these explanations. In a study of FIR line emission from 240 galaxies, \cite{Diaz-Santos2017}, however, attribute the \cii\,  deficit to the contributions from PDRs to the observed \cii\,  emission varying as a function of dust temperature.

In luminous galaxies, LIRGs, and ULIRGs with little AGN activity, the FIR luminosity is thought to arise primarily from dust-embedded high-mass star-forming regions. In the Milky Way, such young high-mass stars form in $\sim$1 pc scale dense molecular clumps usually associated with Infrared Dark Clouds  (e.g., \citealt{Carey1998,Hennebelle2001,Rathborne2006}). It is reasonable then to expect that the global FIR and \cii\,  emission from luminous galaxies could be better understood by systematically characterizing the local \cii\,  and FIR properties of $\sim$pc scale high-mass star-forming clumps in the Milky Way.  Indeed, the typical \cii\, to FIR luminosity ratio for normal galaxies, \lcii$/$\lfir\, $\sim 3 \times 10^{-3}$, is quite similar to the ratio for the Orion star-forming region averaged over a $\sim 1$ pc region, \lcii$/$\lfir\, $\sim 3.8 \times 10^{-3}$ \citep{Goicoechea2015}.  Moreover, toward the high-mass star-forming regions W31C, W49N, and W51, \cite{Gerin2015} find  \lcii$/$\lfir\,  ranging from $2 \times 10^{-4}$ at the FIR continuum peaks to $1.5 \times 10^{-3}$ at their map edges, comparable to the range of values found for galaxies.

Presumably, in external galaxies the emission from hundreds or thousands of such clumps is blended together in the kpc scale telescope beam. If these clumps were to have well-characterized properties, the ratios between \cii\,  and FIR, submm, and molecular line emission could constrain whether these clumps might indeed produce a galaxy's global emission, and, if so, which types of clumps and how many of them are required. Although a few dense clumps have recently been imaged in \cii\, \citep[e.g.,][]{ Beuther2014, Gerin2015,Goicoechea2015,  Pabst2019}, a systematic survey of the \cii\,  emission by dense clumps has yet to be performed.

To characterize the molecular line properties of high-mass star-forming clumps, we conducted the Millimetre Astronomy Legacy Team 90 GHz (MALT90) survey with the 22-meter Mopra Telescope \citep{Foster2011,Foster2013,Jackson2013}.  MALT90 targeted a sample of $\sim3,000$ clumps first identified by ATLASGAL
at 870 \um\,   \citep{Schuller2009} and simultaneously imaged 15 molecular lines near 90 GHz, including \hcop, \hcn, \hnc, \nthp, \cch, and \sio. One of the key benefits of MALT90 is the
measurement of the clumps' velocities \citep{Rathborne2016}, and hence, the determination of their kinematic distances \citep{Whitaker2017}.  \cite{Guzman2015} used {\it Herschel}
and ATLASGAL data toward each clump to determine the dust temperatures and column densities, and along with the MALT90 kinematic distances, we have now calculated the masses, sizes, and
luminosities for an unprecedentedly large sample of high-mass star-forming clumps \citep{Contreras2017}.  This large dataset enables tests of the notion that galaxy-scale FIR emission and the global galactic emission of \cii\,  or other lines proposed as tracers of star formation can be understood as the aggregate of the emission by individual star-forming clumps.

For entire galaxies, \cite{Gao2004} found a linear correlation between \lfir\, and the \hcn\, luminosity ($L_{HCN}$), a tracer of dense molecular gas.  They proposed that this correlation arises because \hcnnt\, directly traces the dense star-forming clumps which have roughly constant star-formation efficiencies (reflected in a relatively constant $L_{HCN}$/\lfir\, ratio for all clumps). To test this idea, \cite{Stephens2016} used the MALT90 sample to extend the work of \cite{Wu2005} to compare the \hcn\, and FIR properties of individual clumps with those of entire galaxies.  Contrary to the proposal of \cite{Gao2004} and the findings of \cite{Wu2005}, \cite{Stephens2016} found  that the Gao--Solomon relation in galaxies could not be satisfactorily explained by the blending of large numbers of high-mass clumps with roughly constant   $L_{HCN}$/\lfir\, ratios in the telescope beam.  Instead, they found that the clumps in the MALT90 sample had a large scatter in the $L_{HCN}$/\lfir\, ratio, and that  far too many high-mass clumps would be required to account for the global HCN and FIR luminosities of galaxies.   They concluded that extended FIR emission from dust associated with more diffuse gas  probably dominates the total  \lfir\, in galaxies and that low-mass star-forming clumps and/or subthermal emission from more diffuse gas could possibly dominate the HCN emission in galaxies. Moreover, the deep mapping studies of Orion A \citep{Kauffmann2017} and Orion B \citep{Pety2017} and the small surveys of \cite{HelferBlitz1997} and \cite{McQuinn2002} all suggest that HCN or CS emission is dominated by subthermally excited, more diffuse gas, rather than themally excited emission from dense clumps.

The explanation of the observed correlation of galactic-scale FIR and \cii\,  line emission thus remains an open and interesting question, especially whether this emission can plausibly be explained as arising from an ensemble of high-mass star-forming clumps.   As a step towards addressing that question, in this communication we report observations of \cii\,  emission from four dense clumps that have been well-characterized through dust continuum  observations from the ATLASGAL \citep{Schuller2009} and Hi-GAL \citep{Molinari2010} surveys, and also through molecular line observations in the 90 GHz regime from the MALT90 Survery \citep{Jackson2013}.

\section{Observations and Analysis}
\label{sec:observations}
The four clumps, labeled by their Galactic $l$ and $b$ coordinates, are \sourceA, \sourceB, \sourceC, and \sourceD\,  (hereafter, G302, G313, G317, and G341). First identified in the ATLASGAL survey   \citep{Schuller2009}, these four sources were selected for this study as compact ($< 60''$), luminous ($>10,000$ \Lsun) star-forming clumps that could be mapped efficiently with good signal-to-noise ratios. Although a larger sample of 50 clumps satisfying these angular size and luminosity criteria was identified from the MALT90 sample, only these four sources were scheduled for SOFIA observations. Known properties of these four sources are summarized in in Table~\ref{table:properties}.

Simultaneous  maps of \cii\, at 158 \um\, and \oi\, at 63 \um\, (hereafter ``\oi$_{63}$'') were made for all four sources with the Field-Imaging Far-Infrared Line Spectrometer (FIFI-LS; \citealt{Fischer2018}) operating on the Stratospheric Observatory For Infrared Astronomy (SOFIA; \citealt{Young2012, Temi2018}).  For G317 and G341 we also mapped \oi\, at 145 \um\,  (hereafter ``\oi$_{145}$'').  The observations were made during SOFIA's New Zealand deployment on 4 and 5 July 2016 at pressure altitudes of 39,000 to 43,000 feet.  The  SOFIA telescope has a diameter of 2.7 meters, yielding a diffraction-limited  FWHM angular resolution of  13\arcsec\, at 158 $\mu$m.  \cii\, and \oi$_{145}$ were observed in the red channel of FIFI-LS, which has a five pixel by five pixel field of view at 12\arcsec\, per pixel.  The spectral resolution of the spectrometer is $R \sim 1200$ for $\lambda$ =  158  $\mu$m, corresponding to a FWHM velocity resolution of $\sim$250 \kms\, and a Gaussian dispersion  of  $\sigma_{\lambda}\sim0.056$ \um\, for a Gaussian model of the spectral response as a function of wavelength.   \oi$_{63}$ data were collected in the blue FIFI-LS channel at second order, with 6\arcsec\, detector pixels and comparable spectral resolution as the red channel.   The observations were conducted using the two-point symmetric chop and nod sky subtraction scheme yielding a ten minute on-source integration for each observation.  Flat-fielding, calibration, telluric corrections, and gridding and smoothing were done in the data-processing pipeline as described in \cite{Fischer2018}. 

We modeled the FIFI-LS spectrum in each pixel as the superposition of  line emission, broadened according to the spectral point spread function of FIFI-LS, and a smoothly varying FIR continuum.  Undersampling introduces strong variations at each end of the $\sim$0.4-1 \um\, spectral range for the spectra at each position.    To produce reliable measurements of both the line and continuum fluxes, we selected a fit window for each source by examining the spectrum averaged over the nine pixels centered on the maximum total emission and picking a channel range that avoided obvious fluctuations.  The selected channel range was then used to analyze the spectrum in each pixel.  The telluric corrections/blanking for the 63 \um\, data were so severe that the \oi$_{63}$ line emission wavelength range was not accessible; we took the average of the fit range amplitudes as the measure of the continuum flux density at 63 \um.  We fit the \cii$_{158}$ and \oi$_{145}$ spectra with a linear continuum  baseline plus a Gaussian spectral line.  To reduce correlations in the linear component parameters, the linear component was described by the intercept at the Gaussian peak line wavelength and a slope; the intercept value is then reported as the continuum level.  

In cases where the fit did not converge (1\%\, of pixels in the 158 \um\, data), the Gaussian amplitude was less than twice its uncertainty (12\%\, of 158 \um\,pixels), or the dispersion $\sigma_\lambda$ of the fit was more than 50\%\, different from the nominal spectral resolution $\sigma_{inst} \sim 0.056$ \um\, (5\%\, of 158 \um\,  pixels), the Gaussian amplitude was set to zero and the continuum level was set to the average of the fit region amplitudes.    The formal relative errors on the continuum and the line flux were in general small; we assign an uncertainty of 15\% to these measurements as an estimate of the absolute calibration uncertainty.

All of the line and continuum fluxes are presented in Table~\ref{table:fifimeas}.

\section{ Results}

\begin{figure}
\begin{center}
\includegraphics[scale=0.35, angle=0]{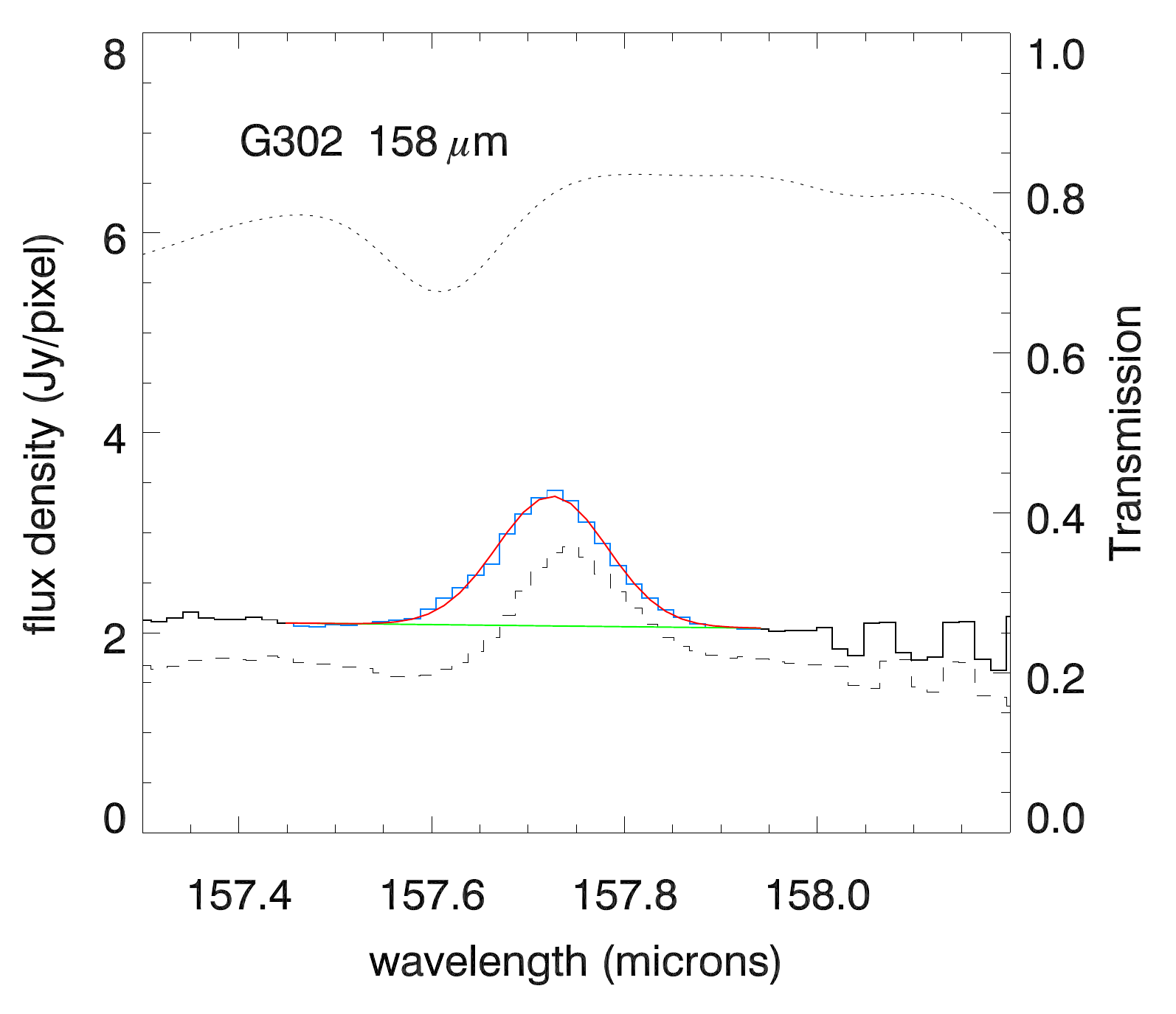} 
\includegraphics[scale=0.33, angle=0]{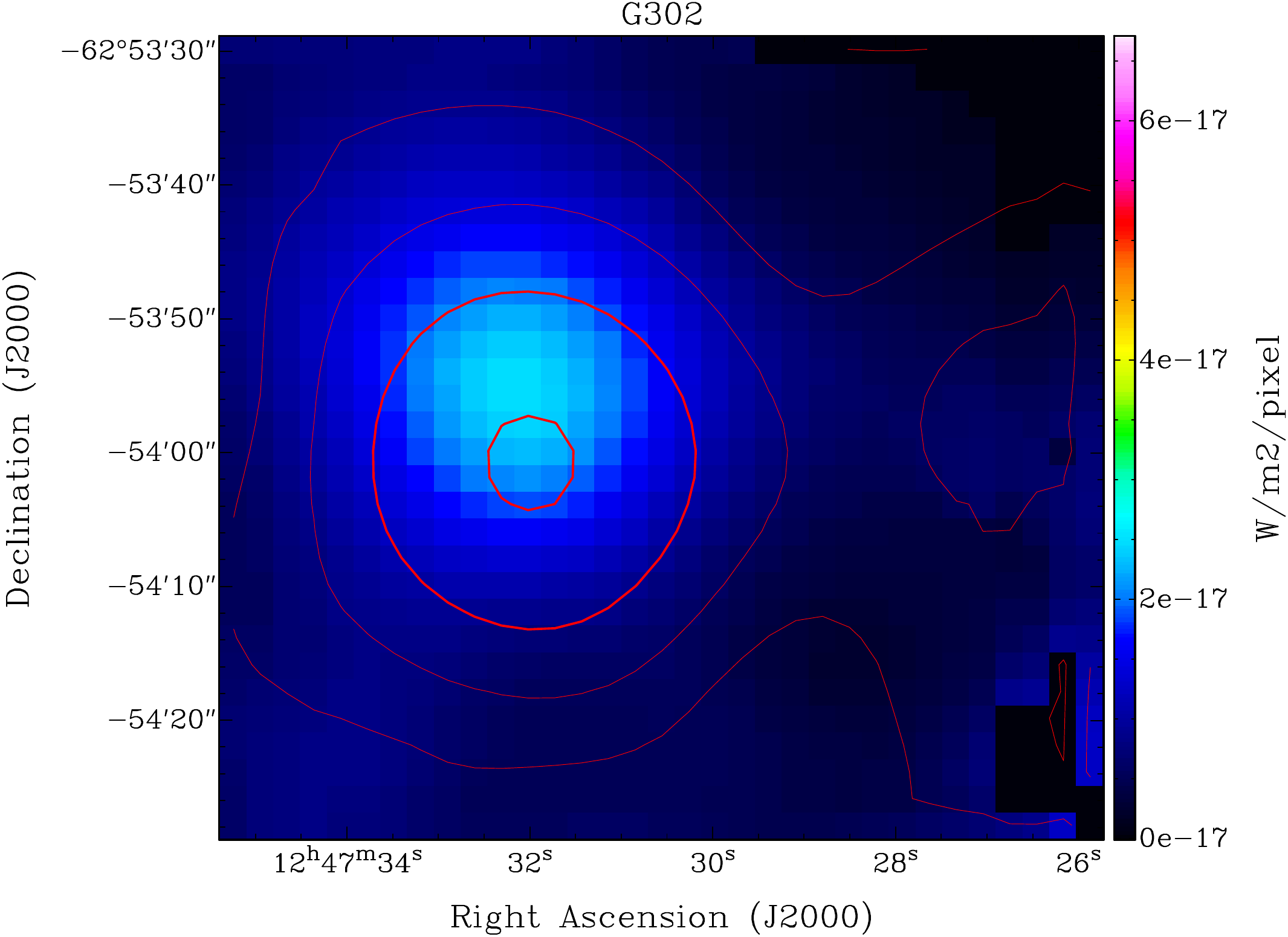} 
\includegraphics[scale=0.35, angle=0]{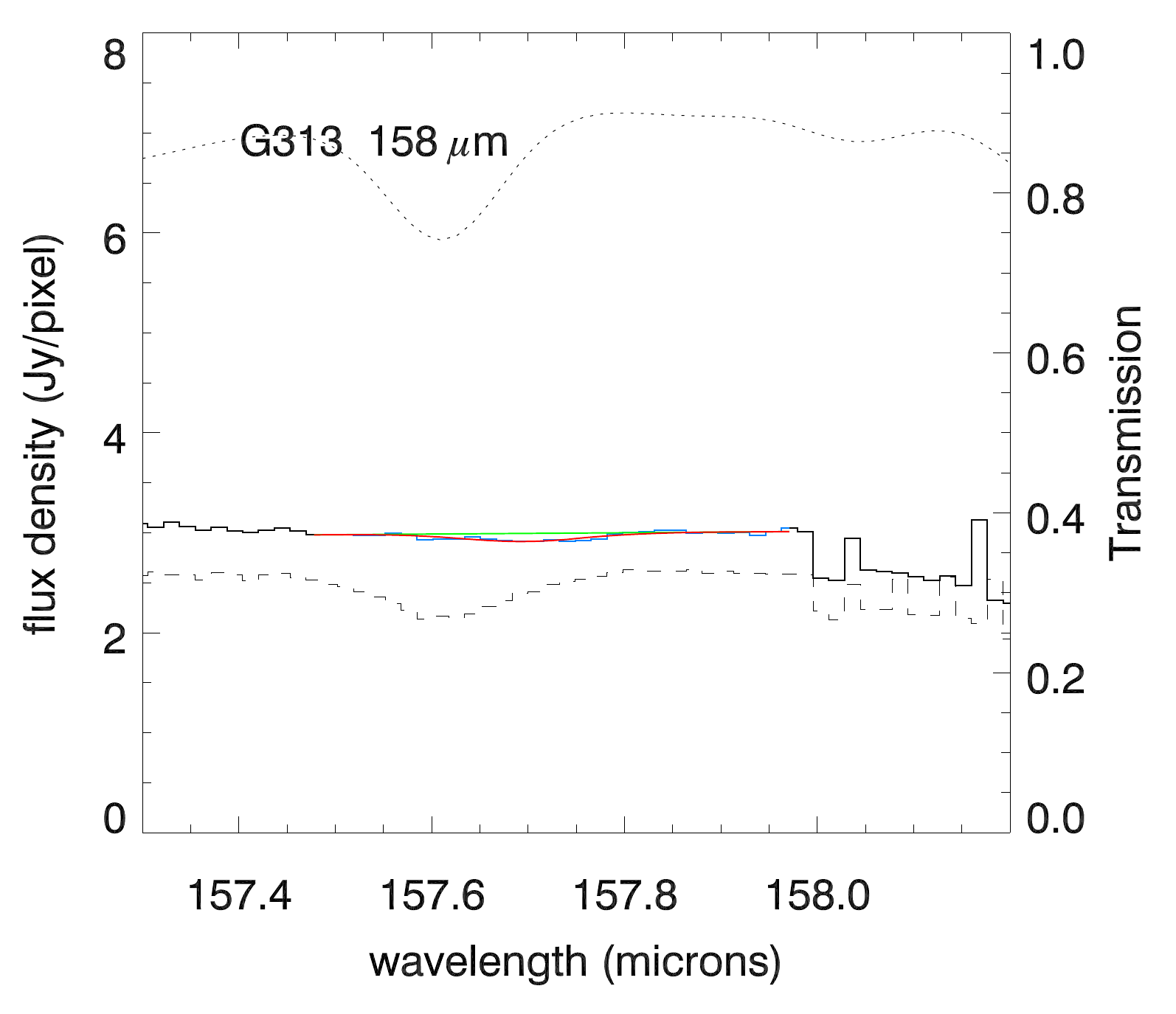} 
\includegraphics[scale=0.33, angle=0]{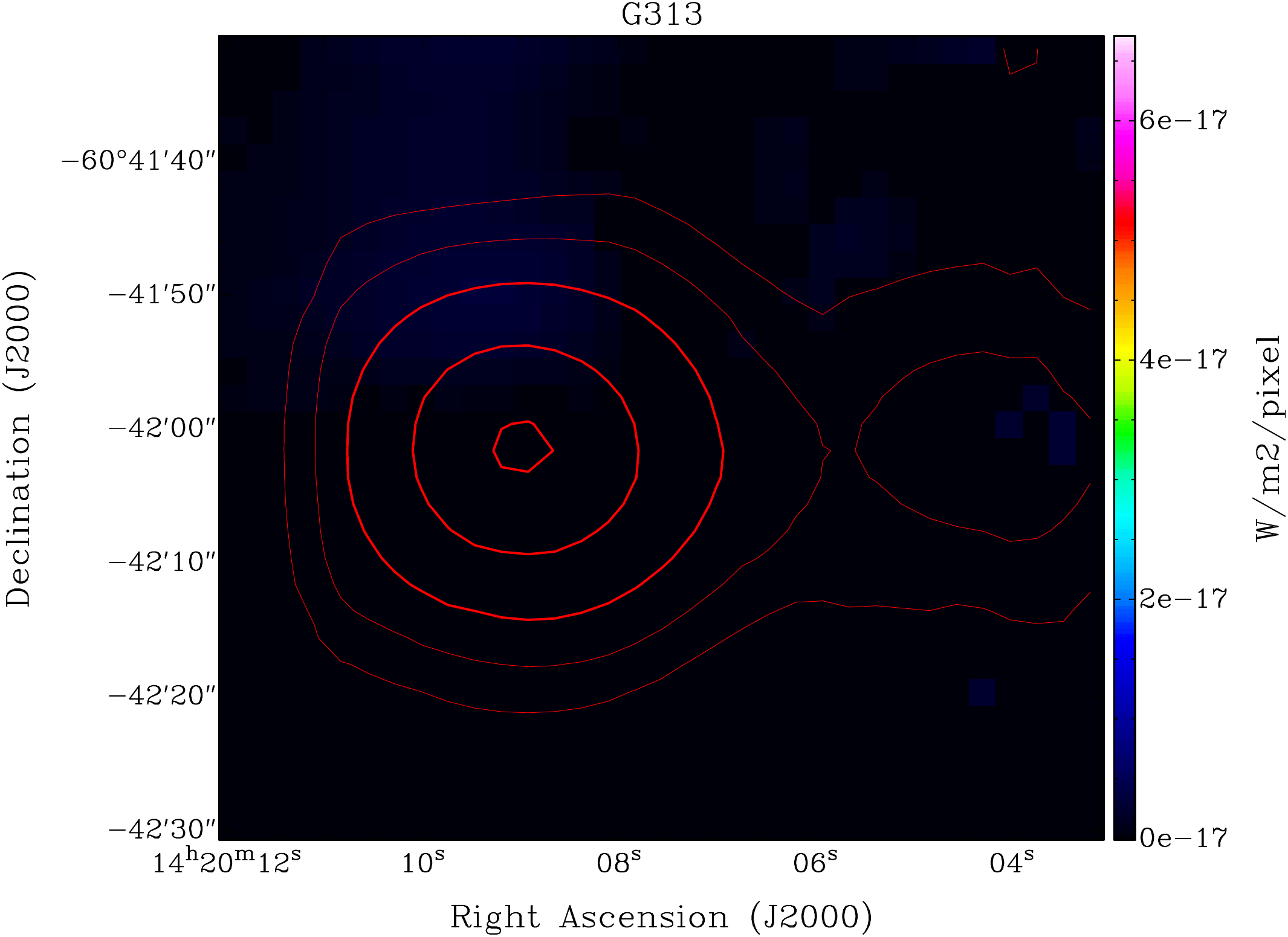} 
\includegraphics[scale=0.35, angle=0]{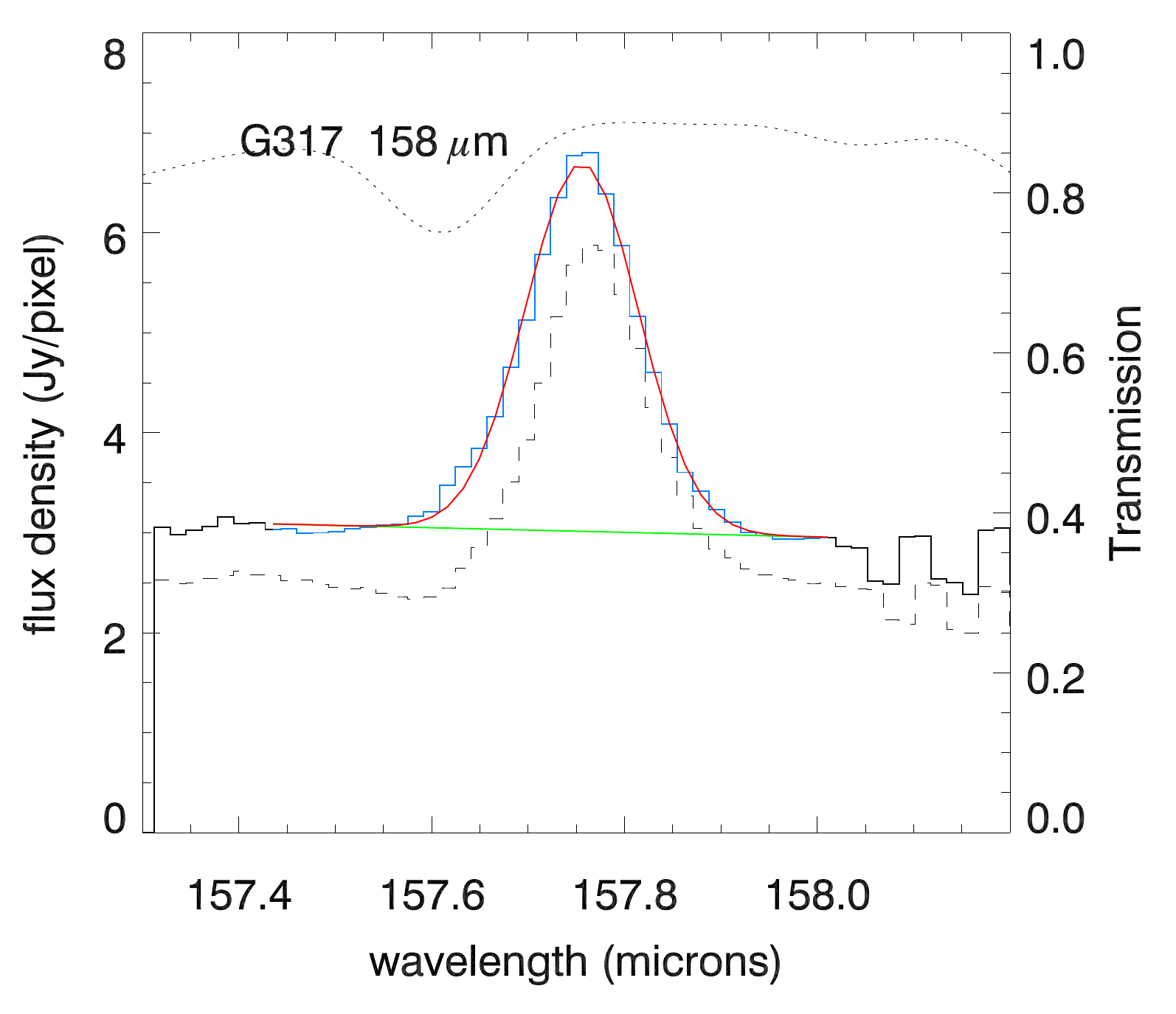} 
\includegraphics[scale=0.33, angle=0]{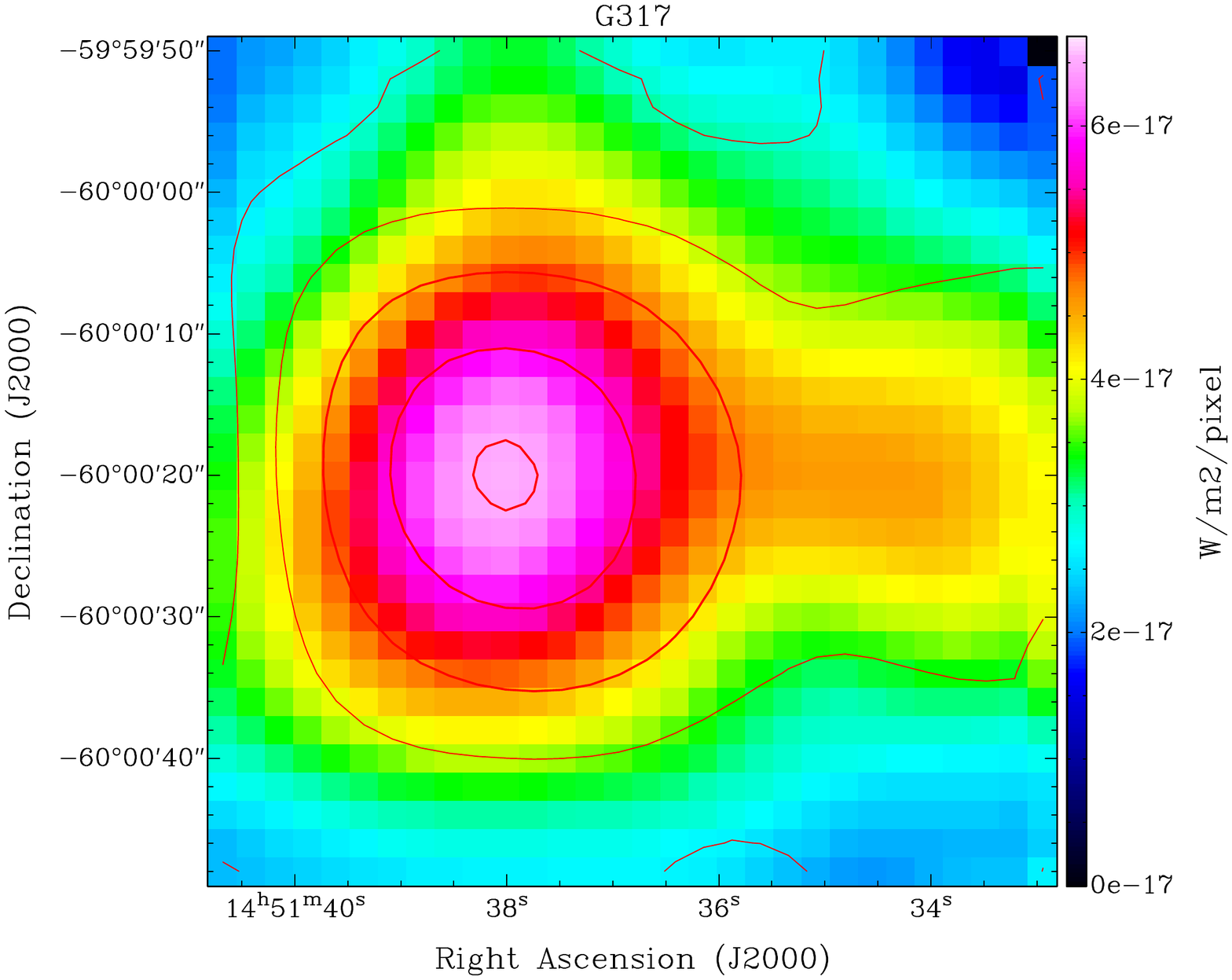} 
\includegraphics[scale=0.35, angle=0]{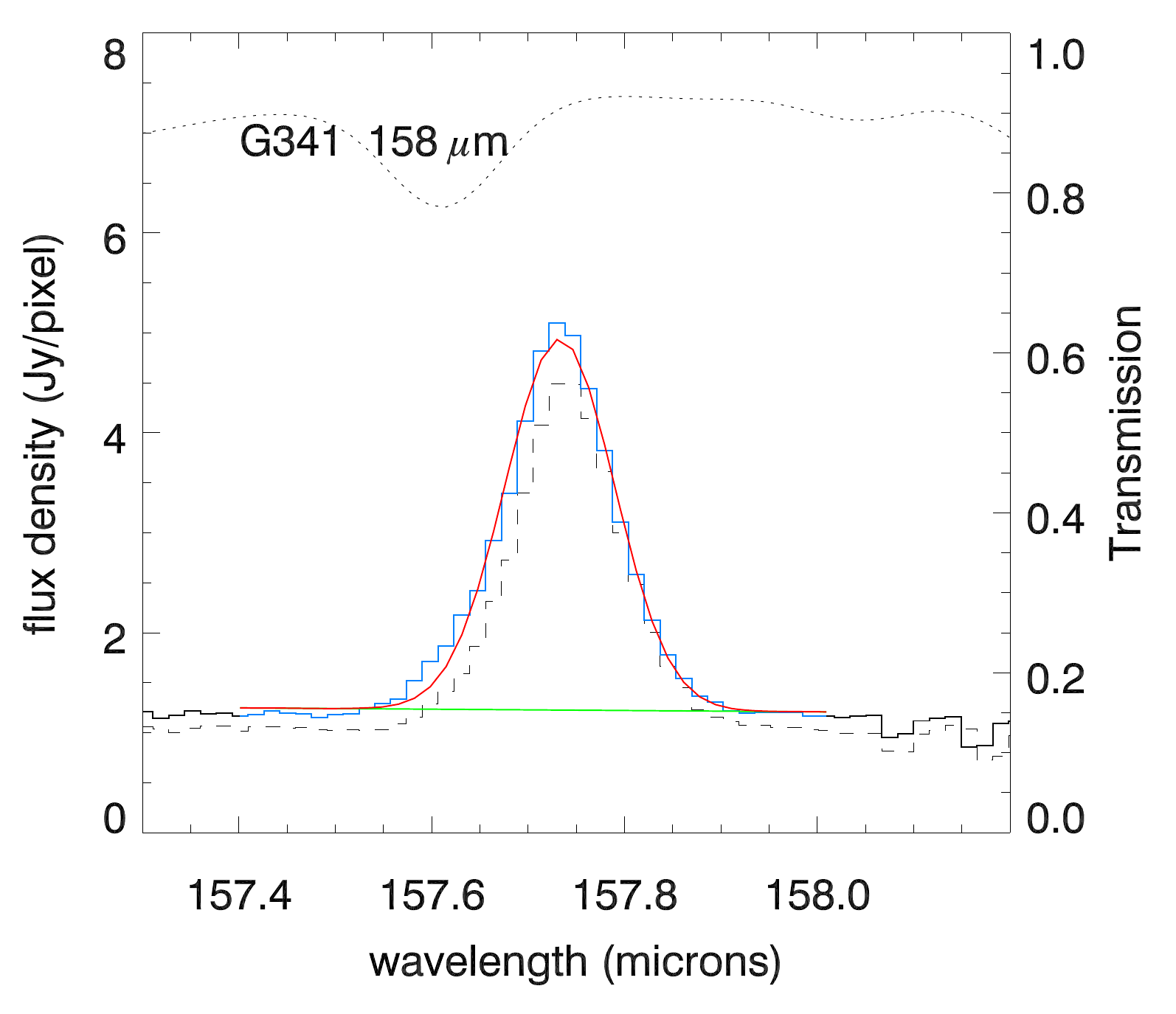} 
\includegraphics[scale=0.33, angle=0]{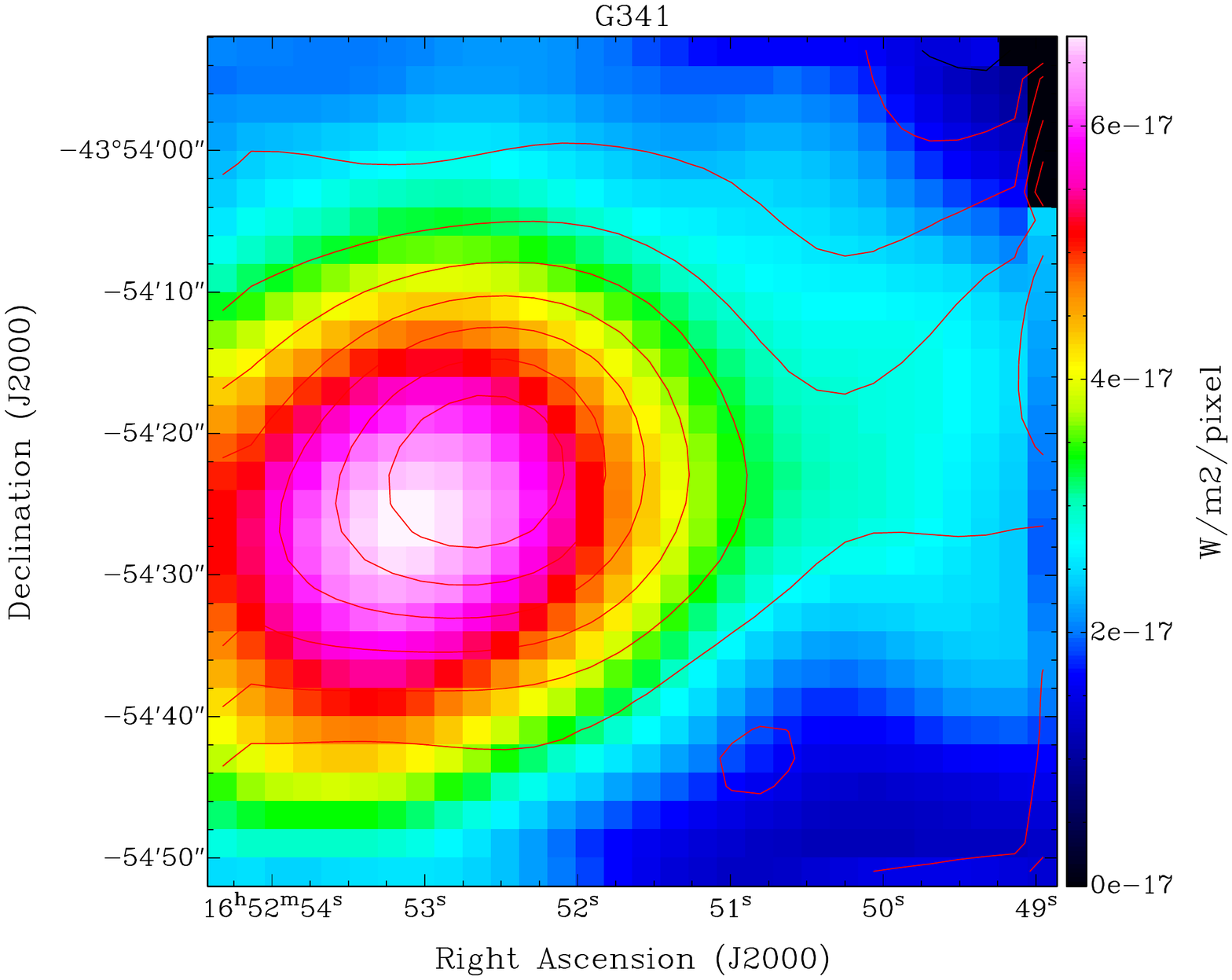} 
\caption{ FIFI-LS \cii\, 158 \um\, spectra (left panels) and images (right panels)  for each of the four target clumps. {\bf  Left panel: \cii\, spectra toward each source.  Right panels: (color) the \cii\ emission; (red contours) the 158 \um\, continuum levels at 0.25, 0.5, 1.0, 2.0, and 3.0 Janskys/pixel.} See text for detailed explanation.}
\label{fig:fig1}
\end{center}
\end{figure}

The 158 \um\, results are shown in  Figure~\ref{fig:fig1}.    The left panels show, to a common scale, the spectrum averaged over a $3 \times 3$ block of pixels centered on the pixel with the maximum total emission. The dashed line indicates the measured intensity uncorrected for telluric absorption, the solid line the intensity after correction for telluric absorption, and the dotted line the nominal atmospheric transmission used for the telluric correction. The linear fit to the continuum is shown in greeen and the Gaussian fit to the line emission is in red. Blue channels indicate the regions used to produce the integrated intensity maps.The peak  continuum emission for all four sources ranges from 1 to 3 Janskys/pixel.  The \cii\,  emission shows greater variation, particularly in its line flux to continuum flux ratio as discussed below.  
 
The right panels in Figure~\ref{fig:fig1} show the integrated \cii\, line intensity; red contours show the continuum levels at 0.25, 0.5, 1.0, 2.0, and 3.0 Janskys/pixel.  Each map shows on the right what appears to be an ``echo'' or a ``ghost'' of the primary emission. These ghosts may be due to instrumental effects such as internal reflections.   The systematic behavior in all four maps makes it improbable that they represent real emission.  As the ``ghost'' images are well separated from the primary source images, their presence does not significantly affect the results.  In subsequent analysis we ignored the continuum and line flux values from the rightmost 20\% of the map.

The panel for G313 in Figure~\ref{fig:fig1} shows no line emission on the continuum peak; instead the \cii\,  spectra show marginal evidence for weak \cii\, absorption against the continuum emission to the north of the continuum peak. 

\begin{figure}
\begin{center}
\includegraphics[scale=0.39, angle=0]{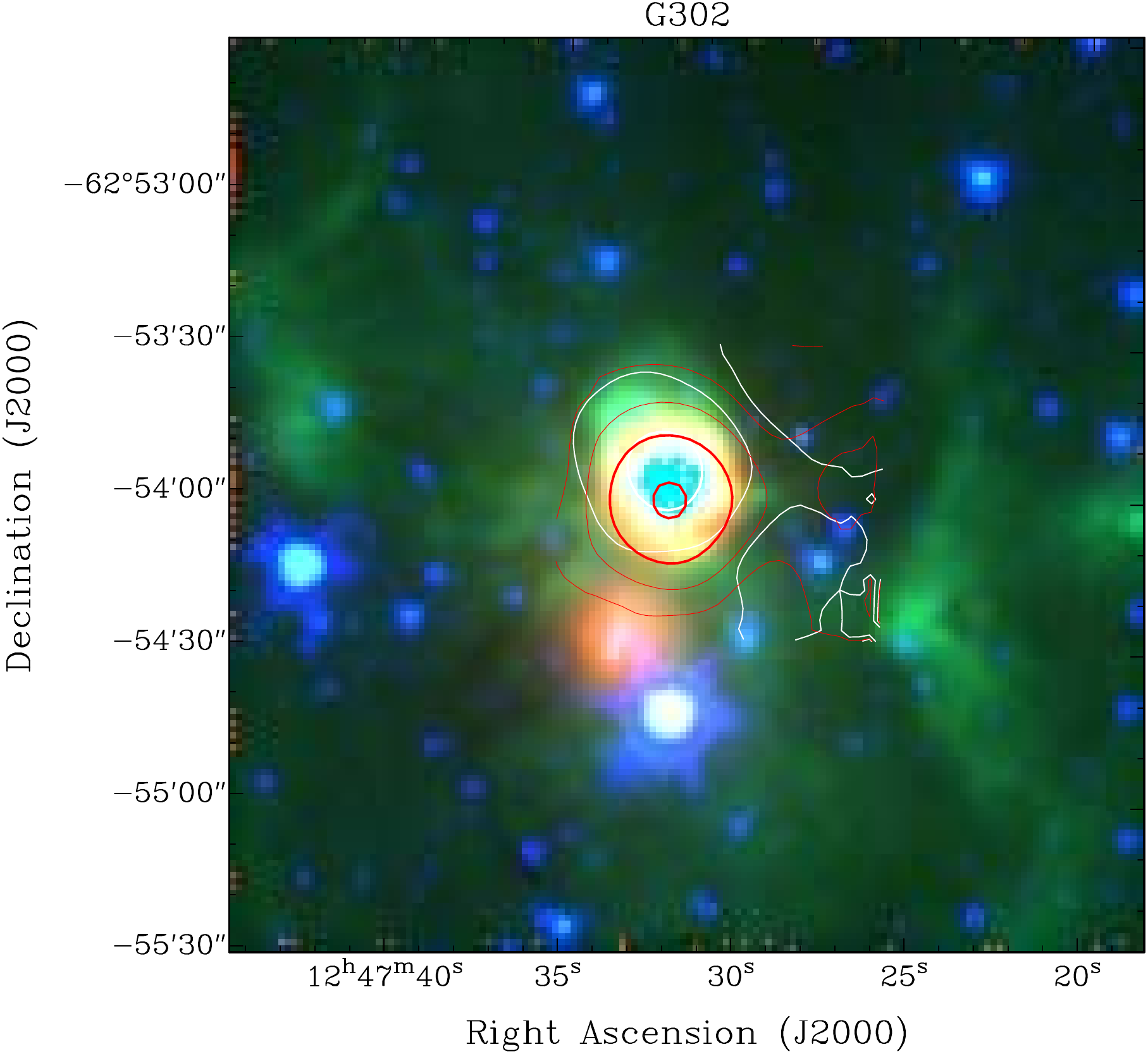}
\includegraphics[scale=0.48, angle=0]{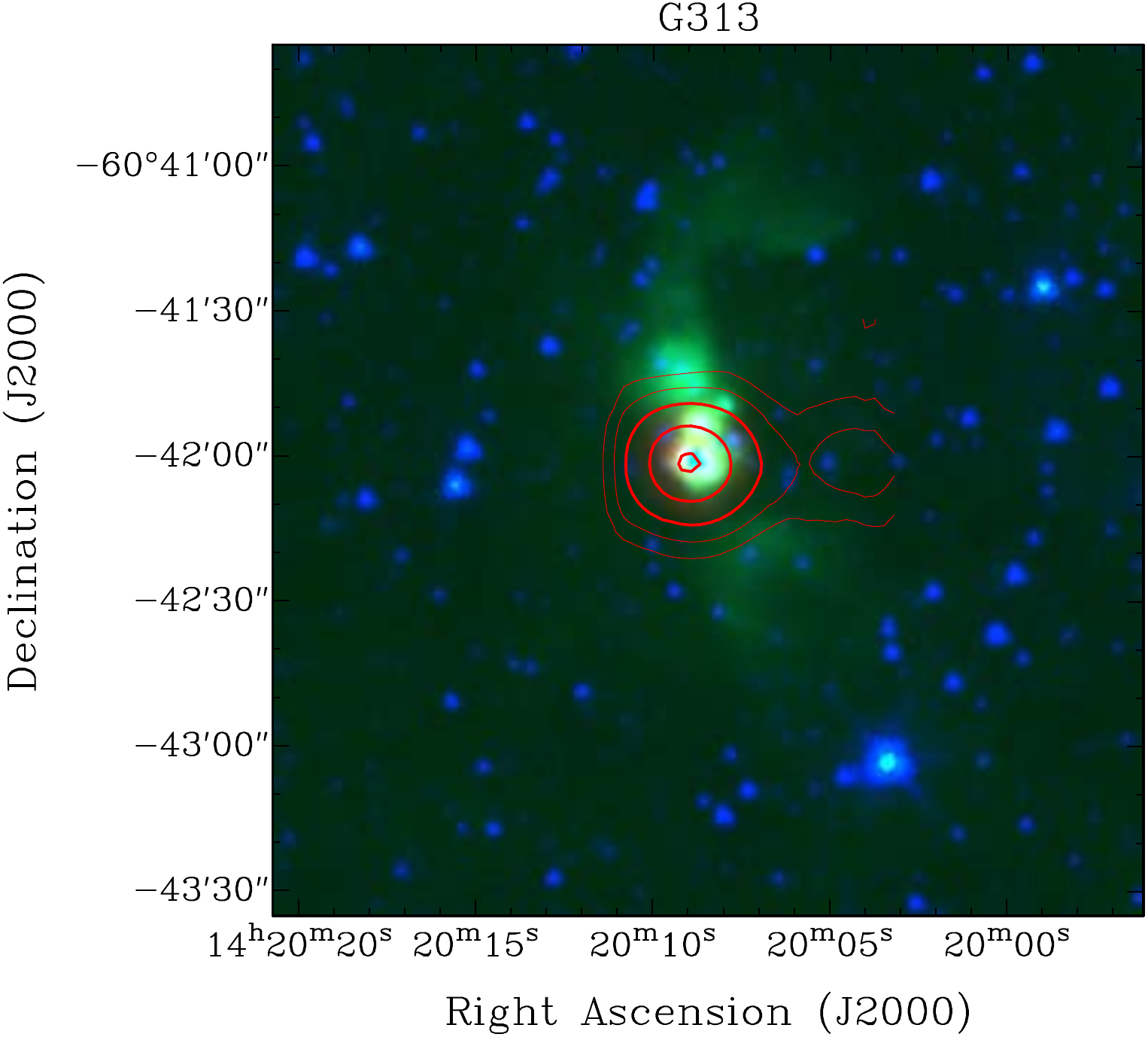}
\includegraphics[scale=0.48, angle=0]{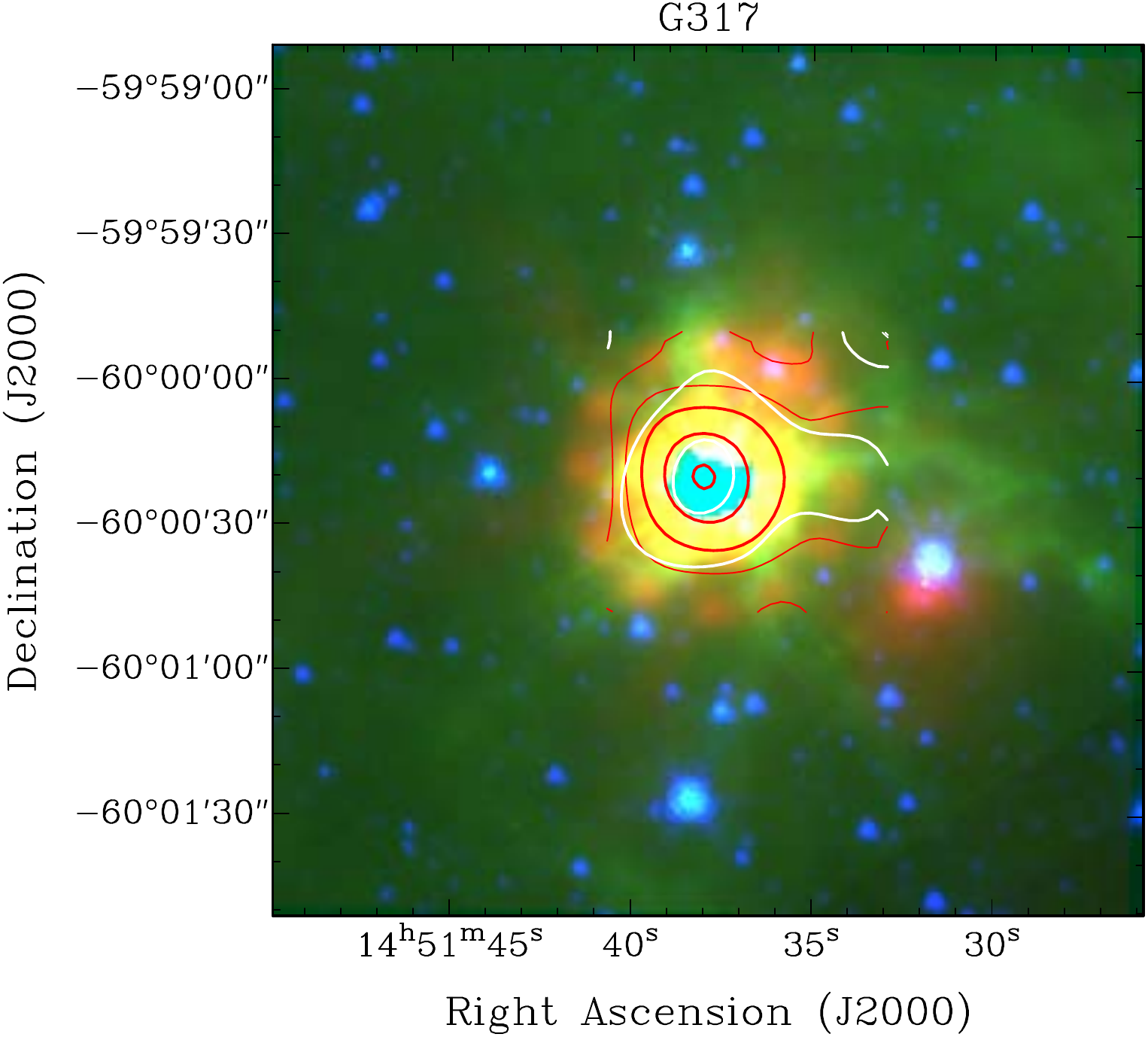}
\includegraphics[scale=0.48, angle=0]{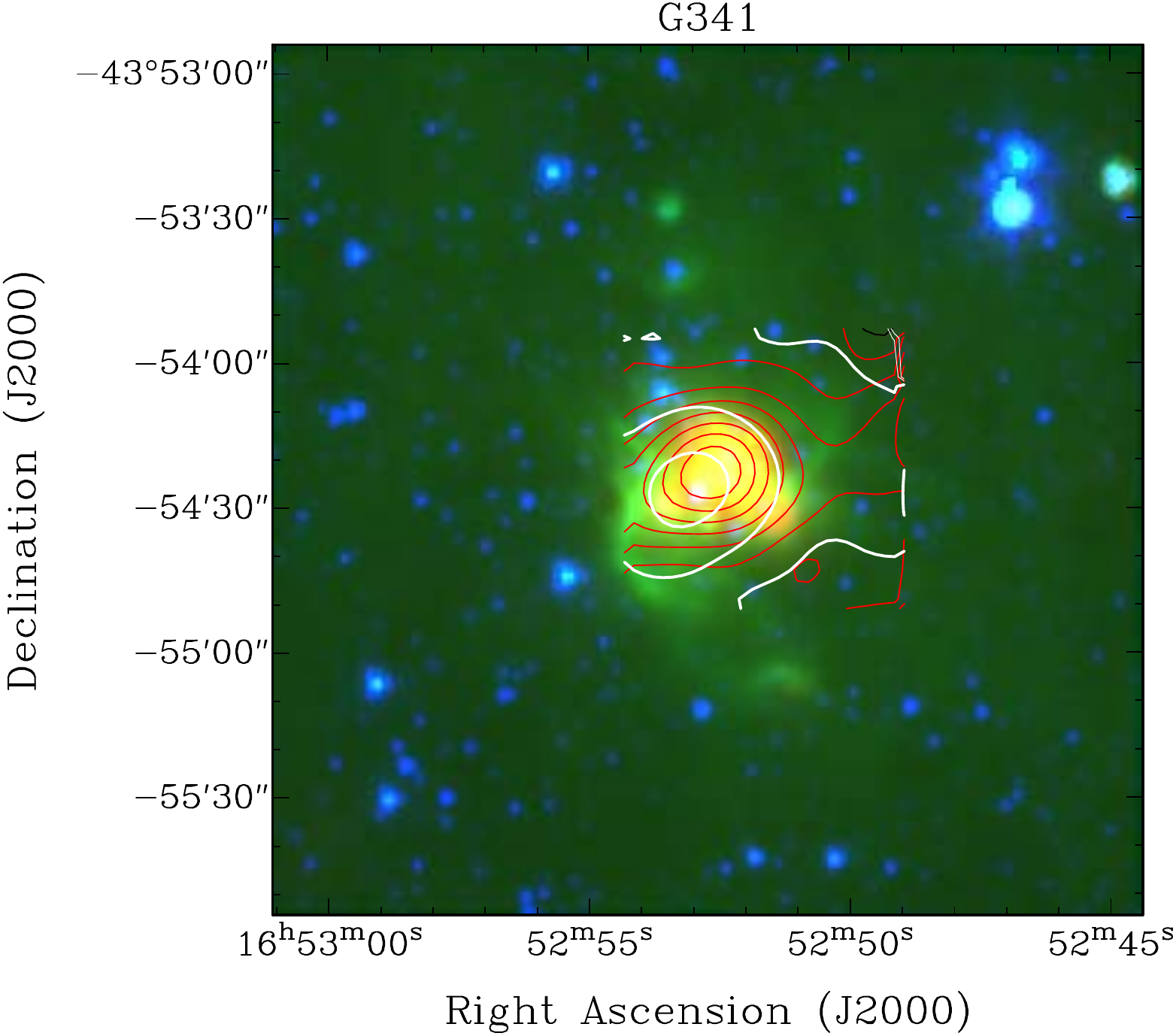}
\caption{Spitzer images with 24 \um\, in red, 8 \um\, in green, and 3.6 \um\, in blue. The red contours show the 158 \um\, continuum with levels the same as in Figure \ref{fig:fig1}; the white contours show the  the \cii\, 158 \um\, line flux  at 0.5, 1.0, 2.0, 4.0, and 6.0 $\times 10^{-17}$ watts/m$^2$/pixel.}
\label{fig:fig2}
\end{center}
\end{figure}

\begin{figure}
\begin{center}
\includegraphics[scale=0.34, angle=0]{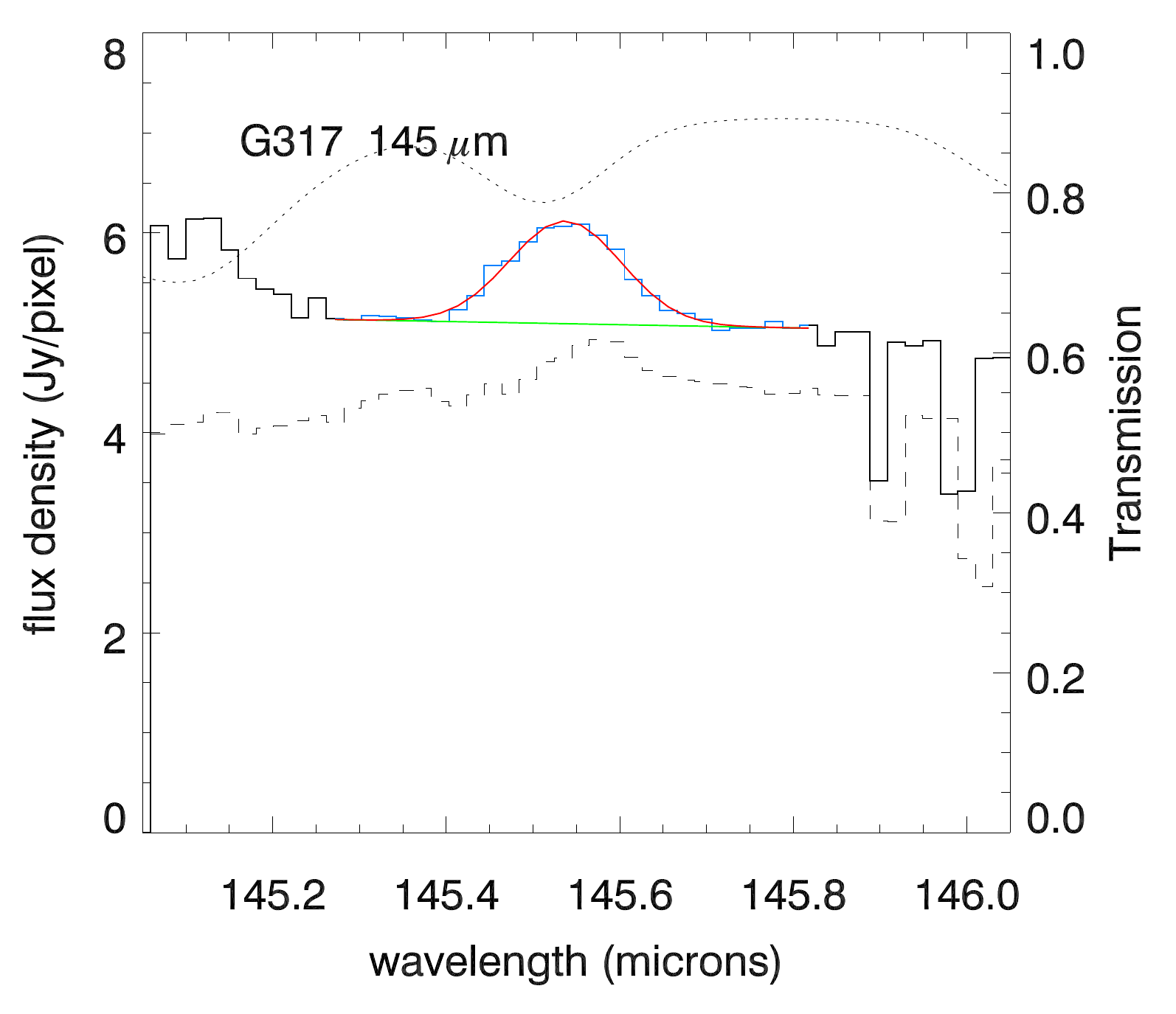} 
\includegraphics[scale=0.34, angle=0]{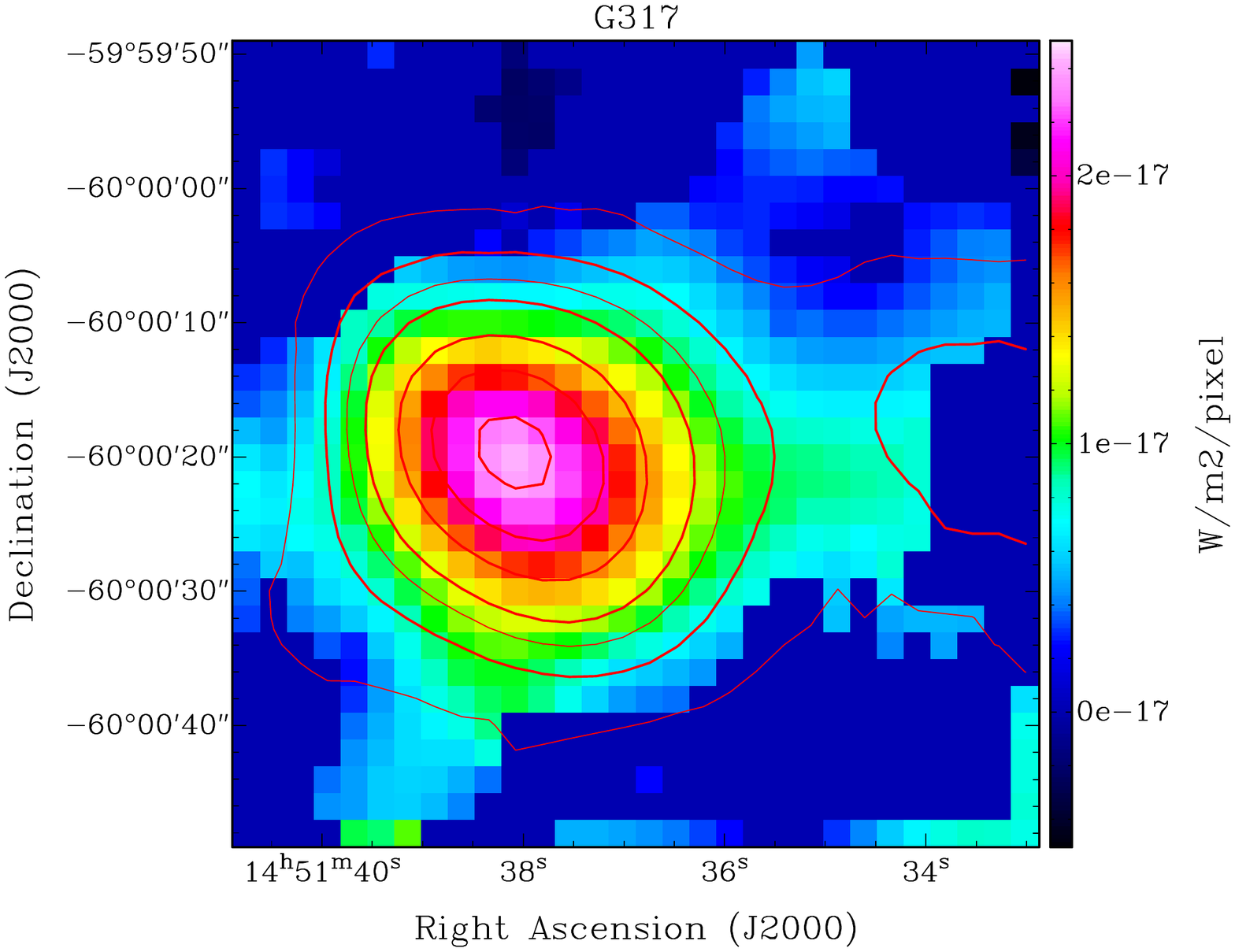} 
\includegraphics[scale=0.34, angle=0]{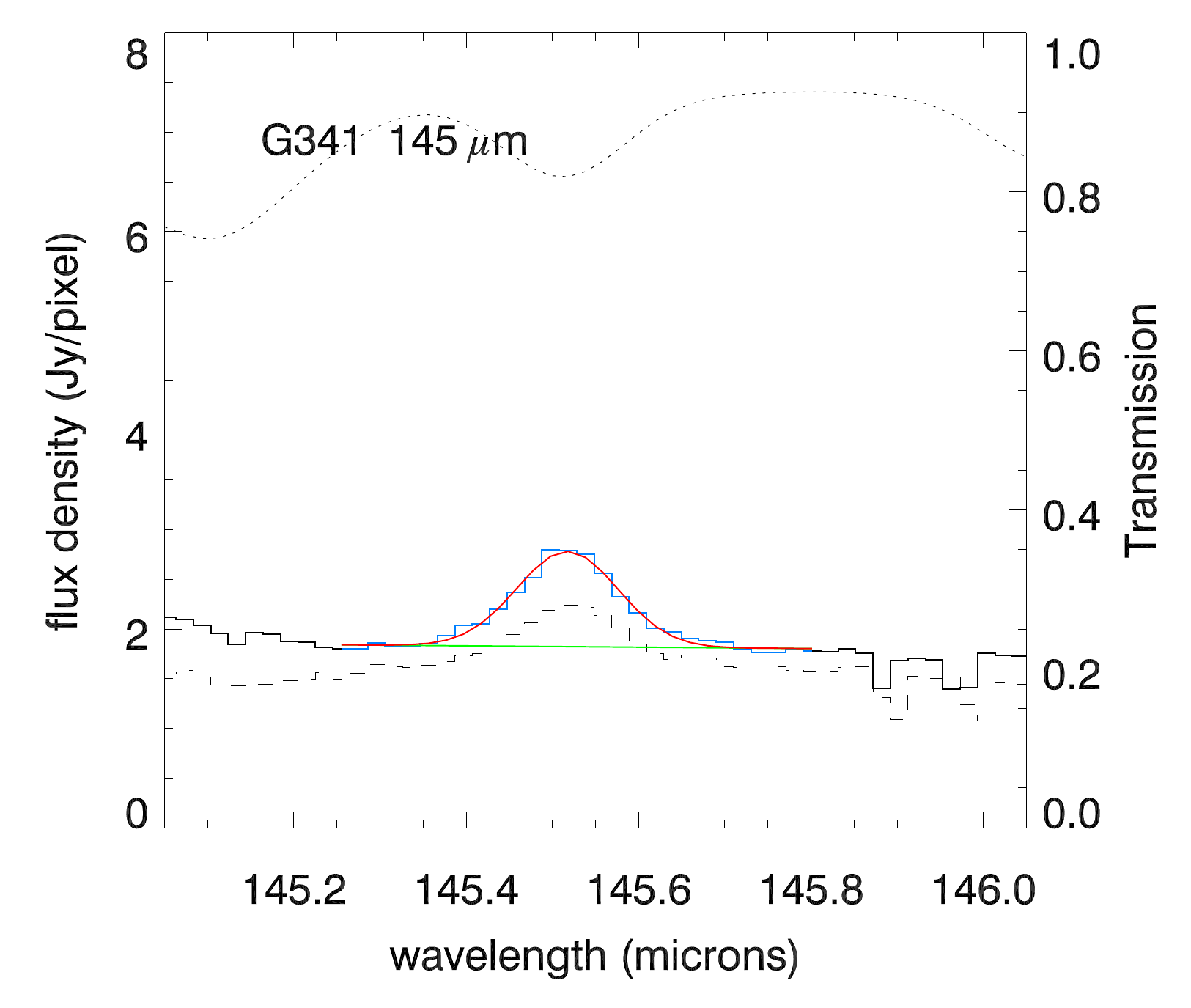} 
\includegraphics[scale=0.34, angle=0]{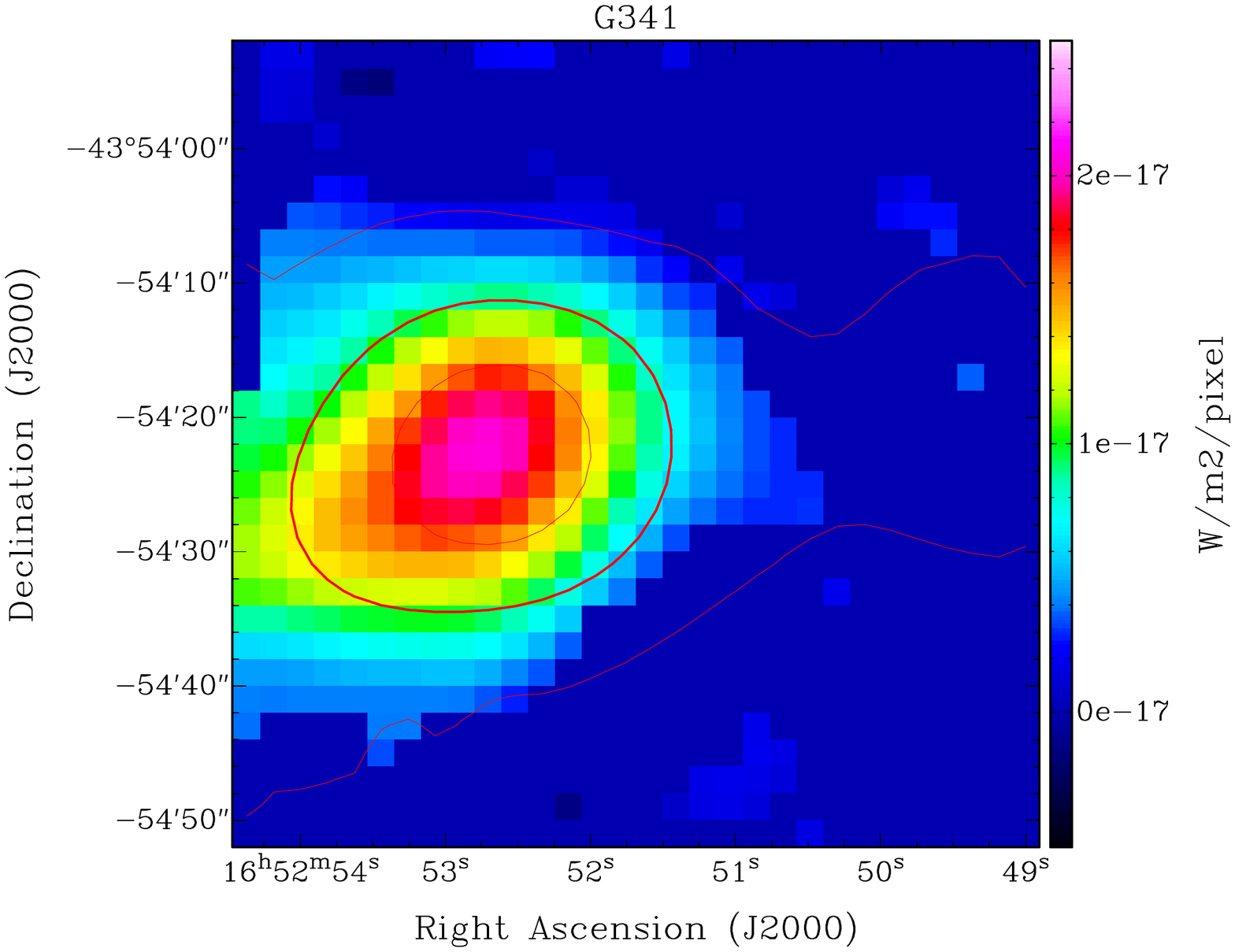} 
\caption{ FIFI-LS 145 \um\, spectra (left panels) and images (right panels) are shown for G317 and G341.  The left panels show, to a common scale,  the spectrum averaged over a $3 \times 3$ block of pixels centered on the pixel with the maximum total emission.  The dashed line indicates the measured intensity uncorrected for telluric absorption; the solid line the intentisity after correction for telluric absorption, and the dotted  line the nominal atmospheric transmission used for the telluric correction.  The linear fit to the continuum is shown in greeen and the Gaussian fit to the line emission is in red.  Blue channels indicate the regions used to produce the integrated intensity maps. The image in the right panel shows the \oi$_{145}$ flux; red contours show the continuum at 0.5, 1.0, 1.5, 2.0, 3.0, 4.0, and 5.0 Jy/pixel. }
\label{fig:fig3}
\end{center}
\end{figure}

\begin{figure}
\begin{center}
\includegraphics[scale=0.38, angle=0]{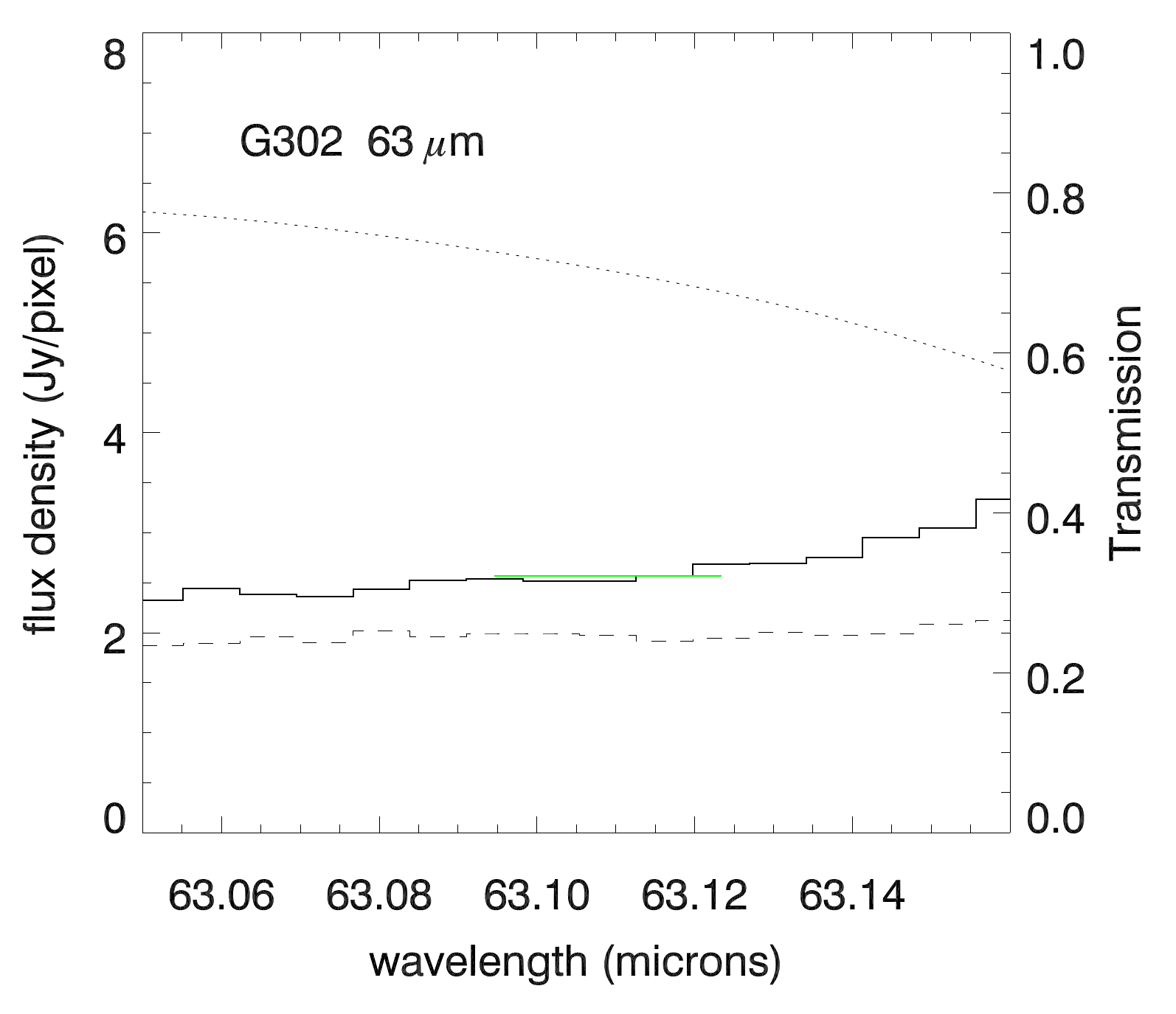} 
\includegraphics[scale=0.3, angle=0]{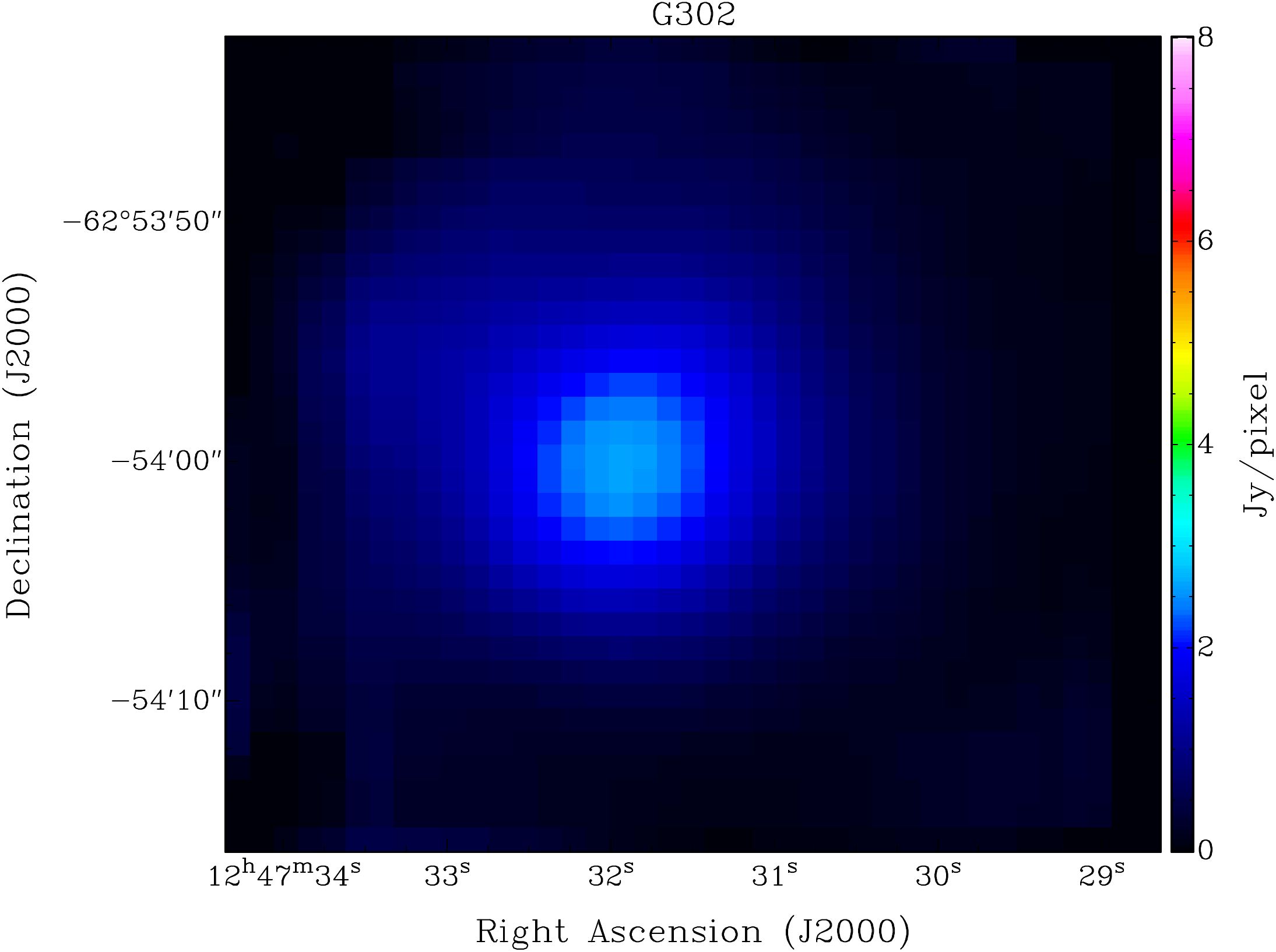} 
\includegraphics[scale=0.38, angle=0]{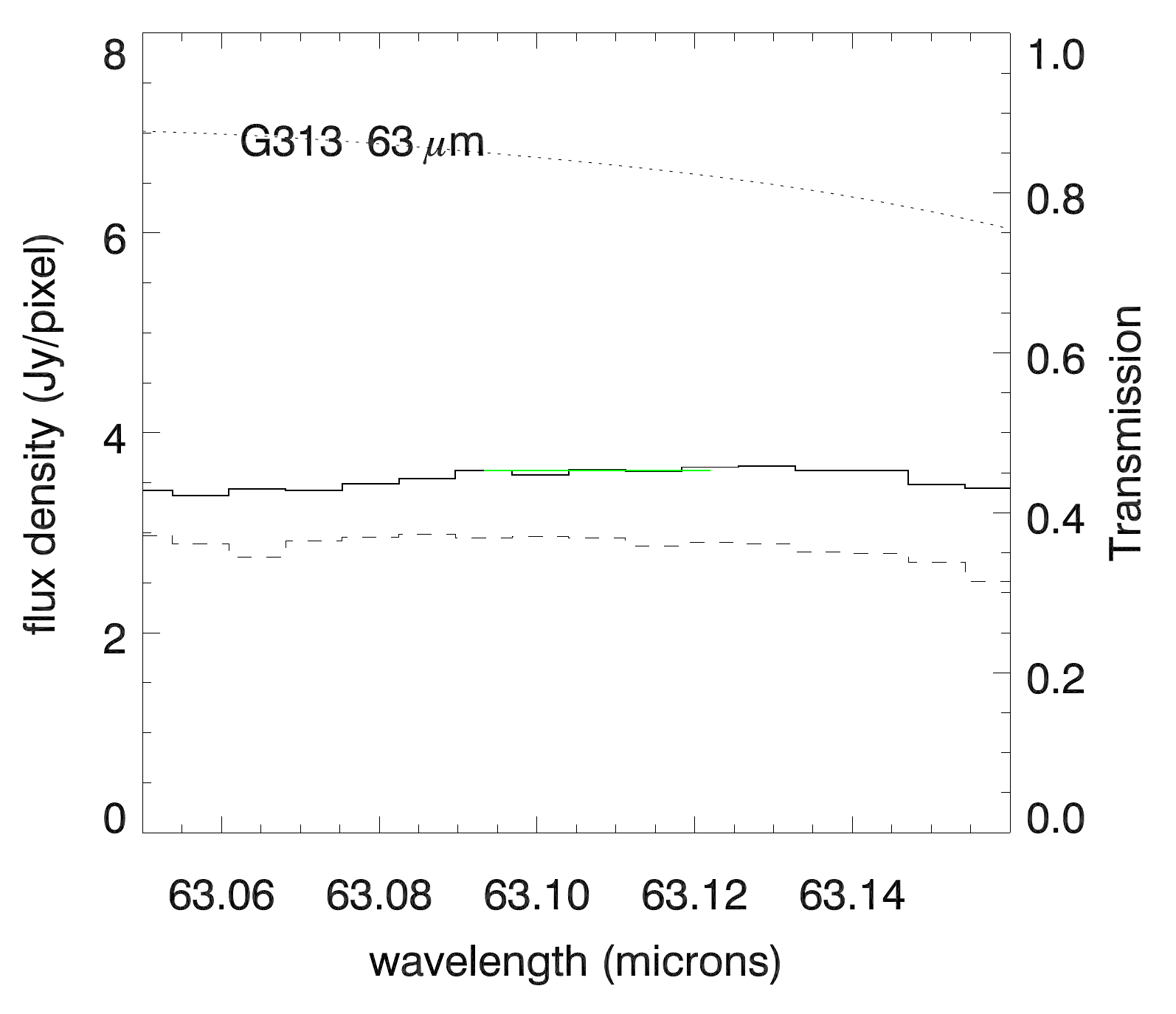} 
\includegraphics[scale=0.3, angle=0]{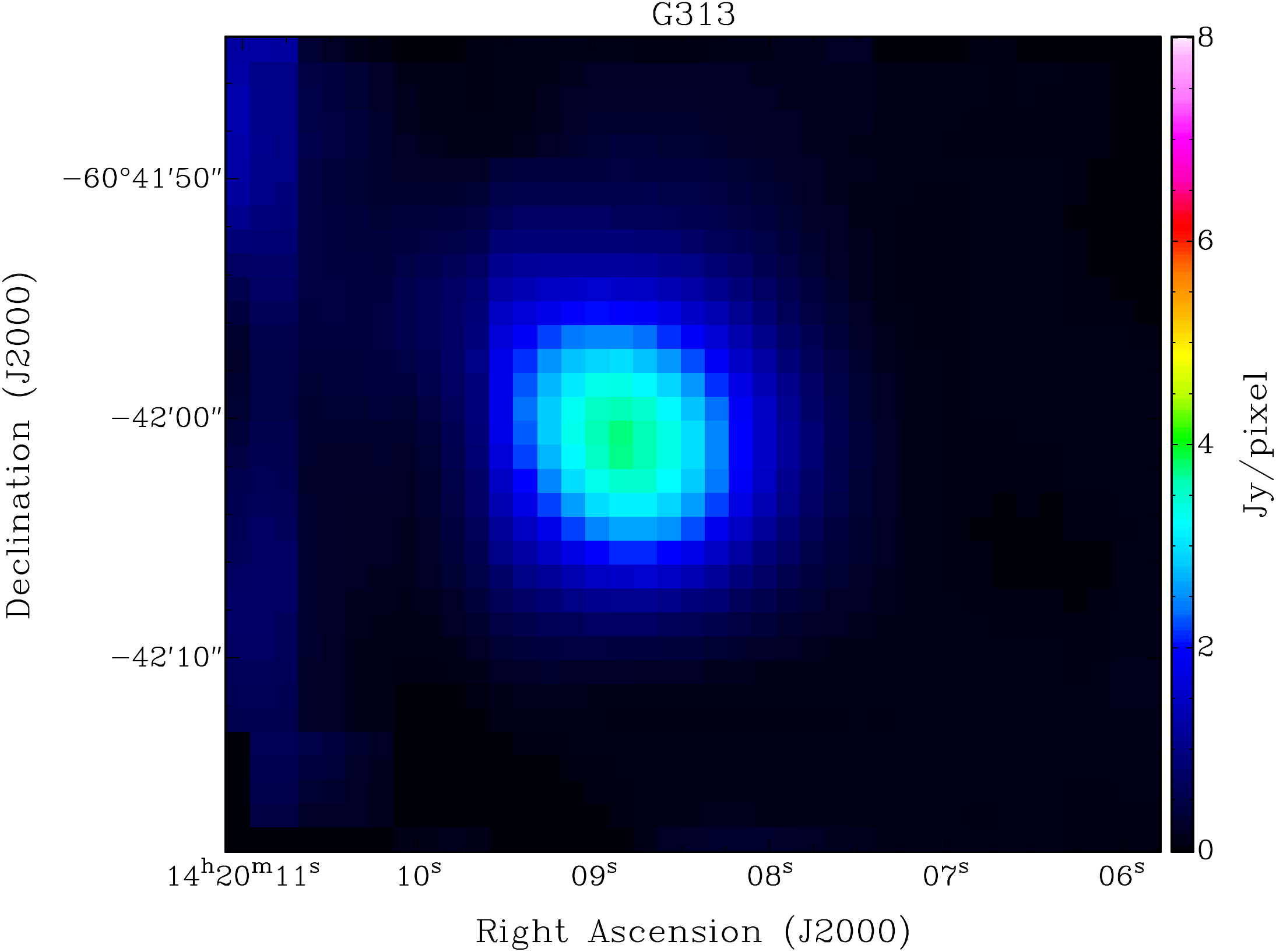} 
\includegraphics[scale=0.38, angle=0]{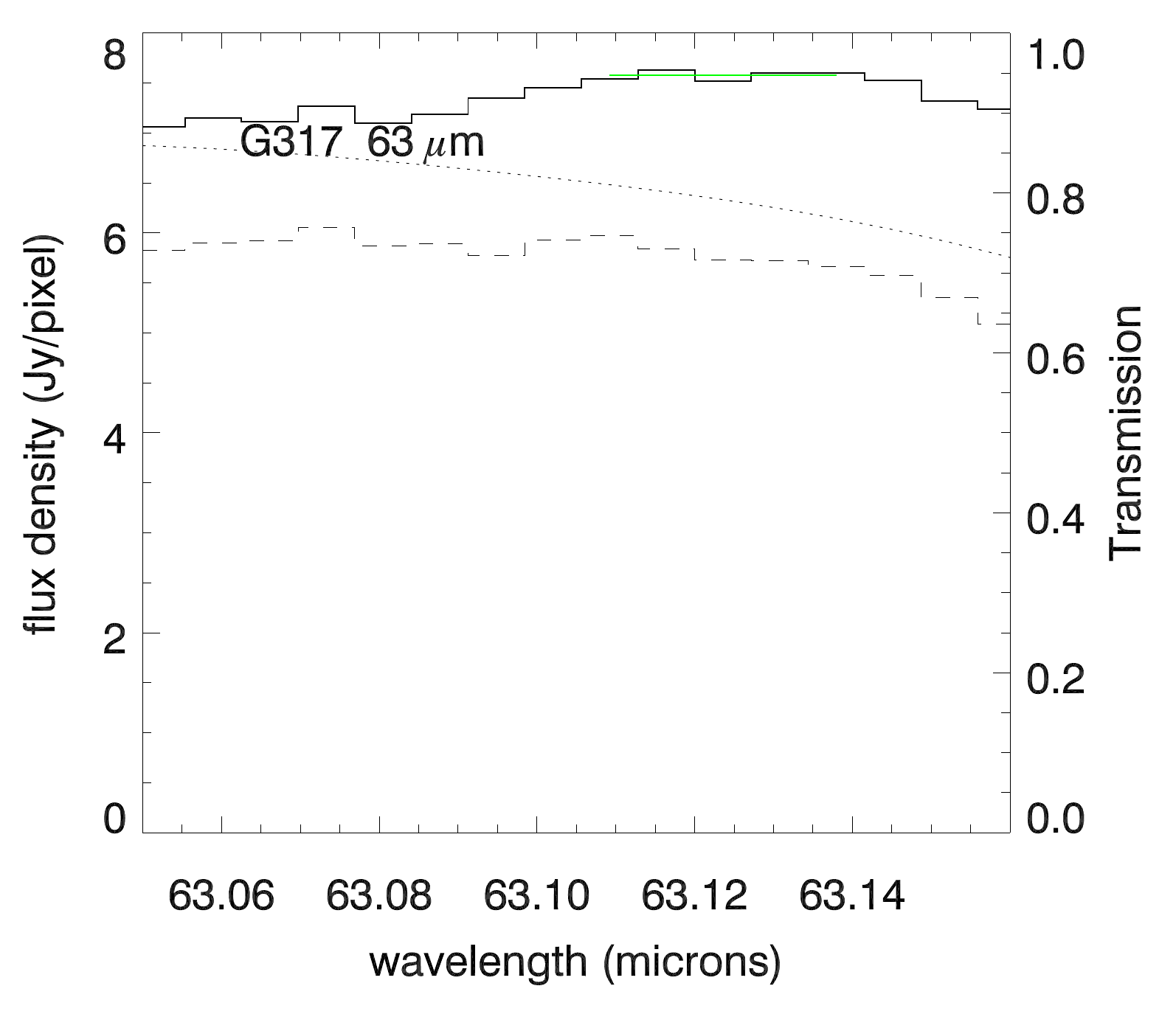} 
\includegraphics[scale=0.3, angle=0]{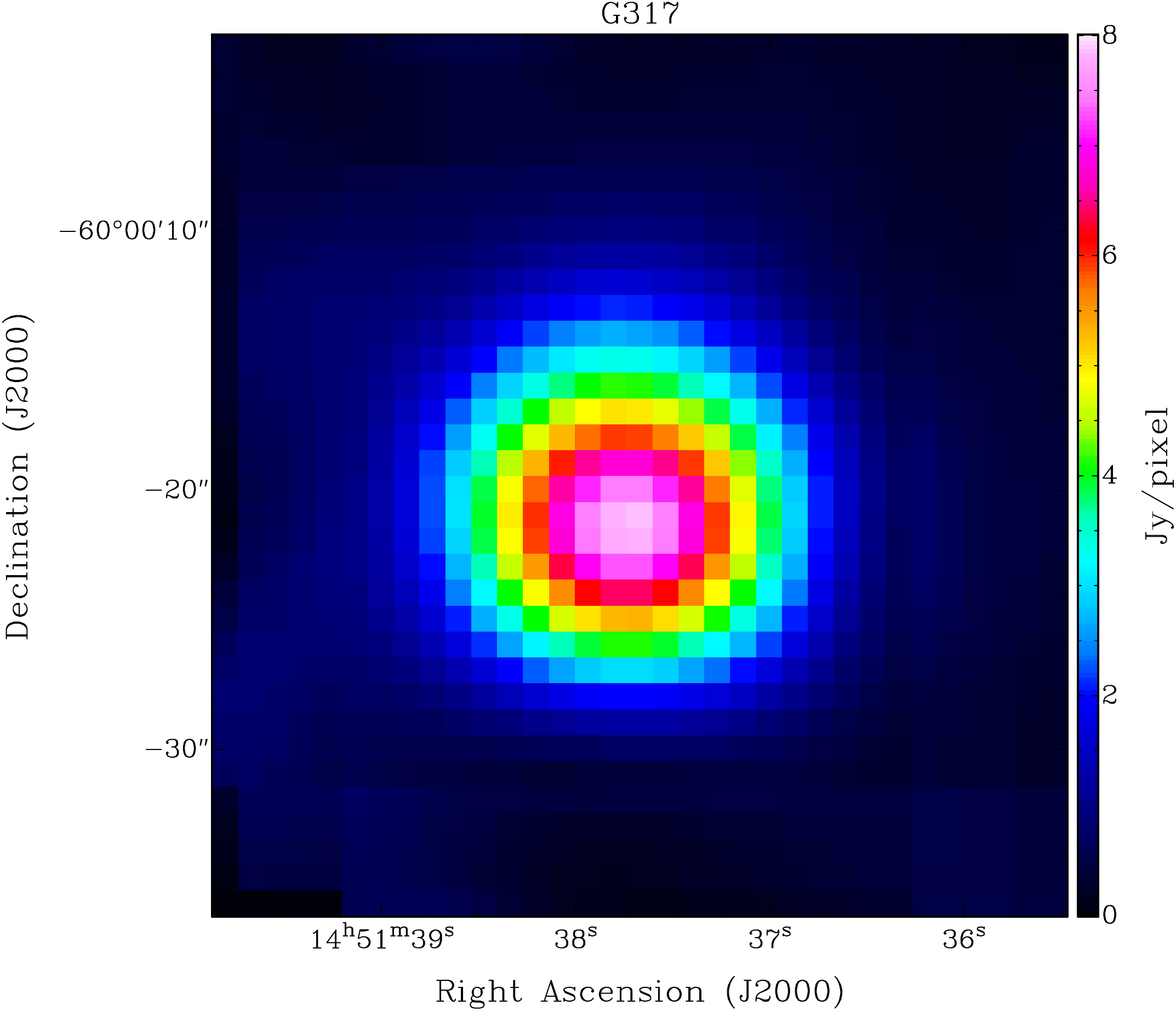} 
\includegraphics[scale=0.38, angle=0]{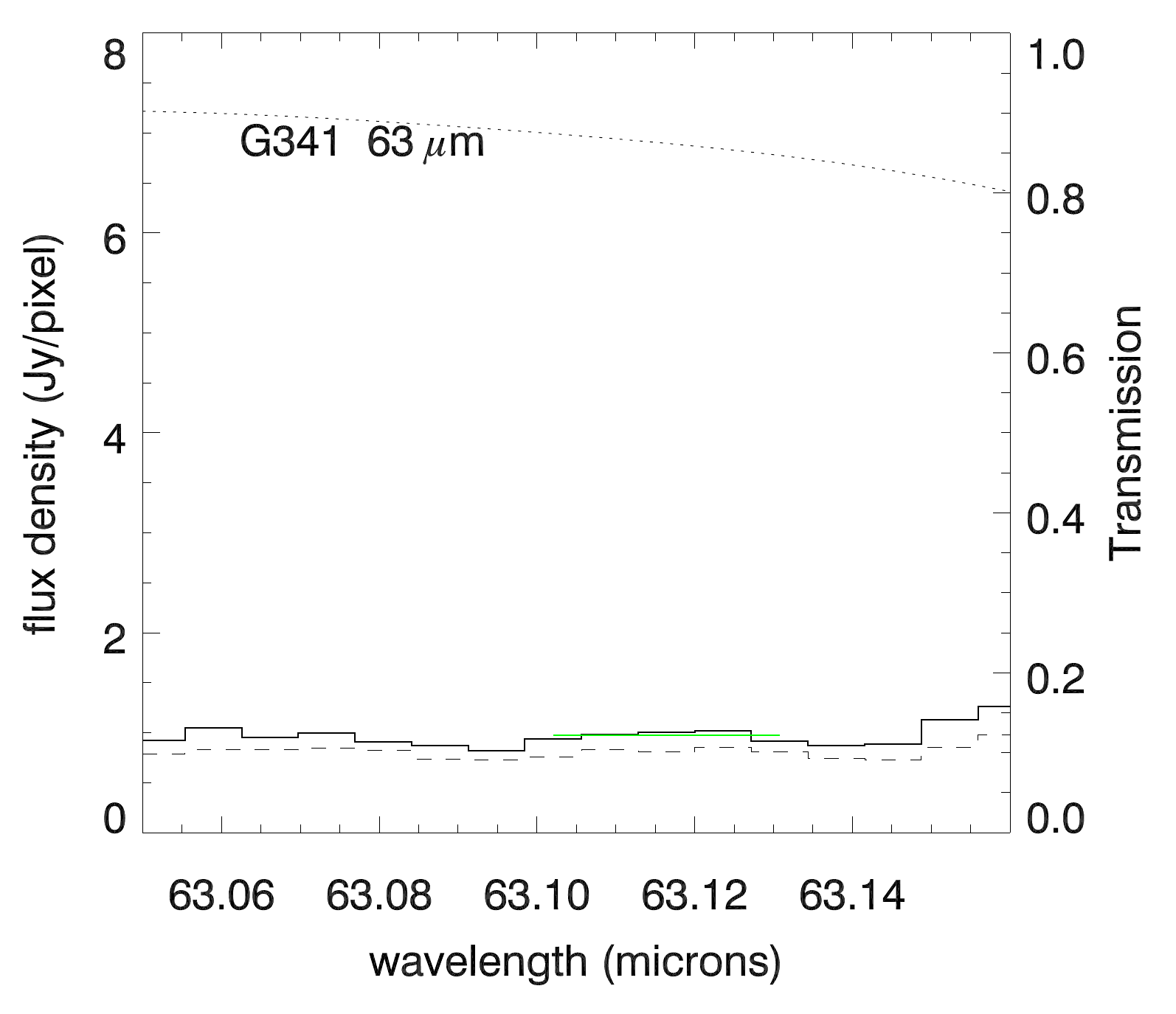} 
\includegraphics[scale=0.3, angle=0]{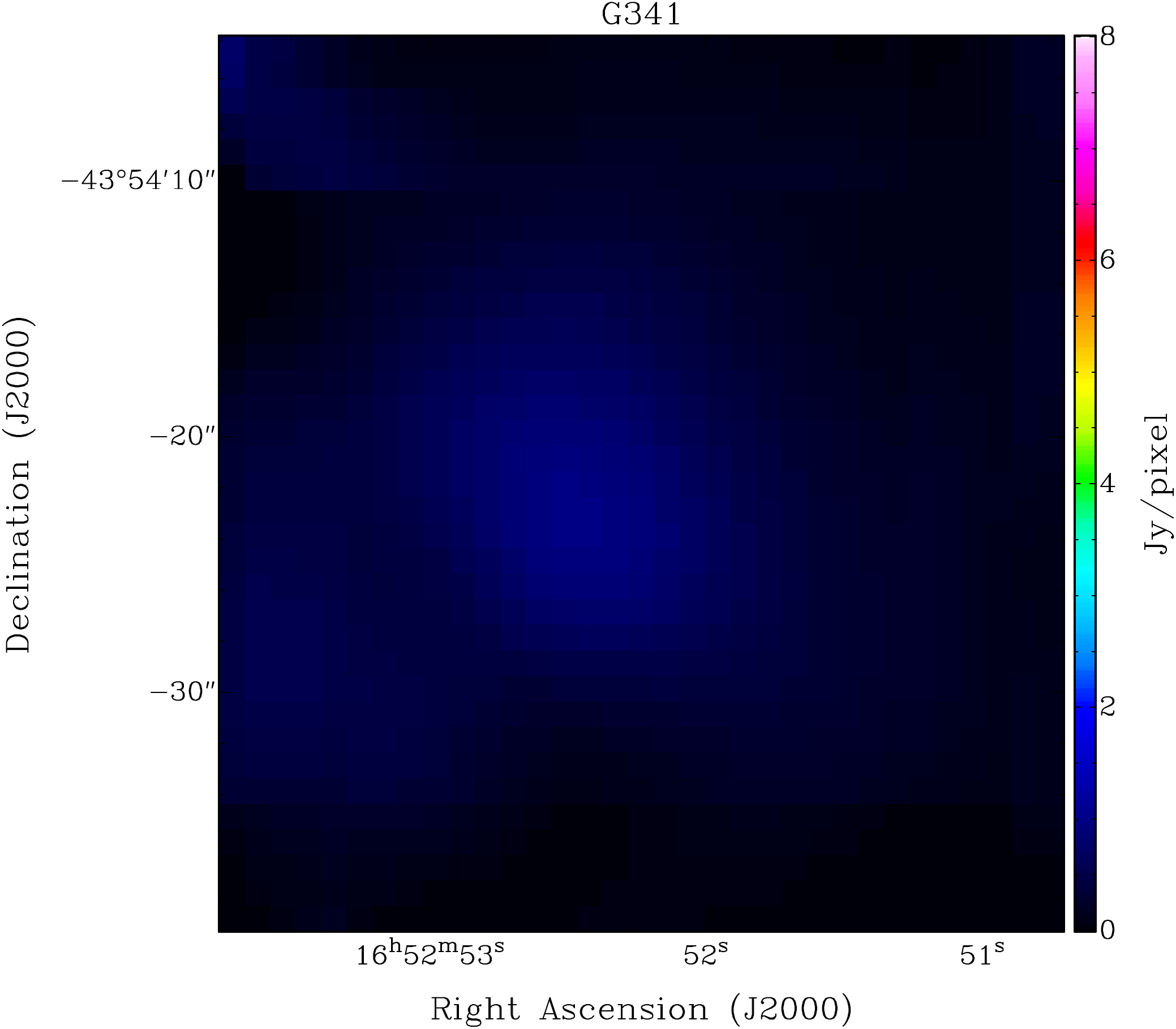} 
\caption{ FIFI-LS results at 63 \um.   The left panels show, to a common scale,  the spectrum averaged over a 3X3 block of pixels centered on the pixel with maximum integrated emission. The green line marks the channels used for continuum measurement.  The image in the right panel shows the  continuum map in Janskys/pixel to a common scale. }
\label{fig:fig4}
\end{center}
\end{figure}

Figure~\ref{fig:fig2} shows 3 arcmin $\times$ 3 arcmin {\it Spitzer} IRAC/MIPS images of each source, with 24 \um\, in red, 8 \um\, in green, and 3.6 \um\, in blue.  The contours for the 158 \um\, continuum and the \cii\, line flux 1 arcmin $\times$ 1 arcmin image are overlaid in red and white respectively.  The white \cii\,  line flux contours tend to follow the green 8 \um\, emission in the Spitzer image.  This is expected since the 8 \um\, emission is associated with PAH fluorescence excited by UV radiation (e.g., \citealt{Allamandola1989}), and UV radiation also ionizes atomic carbon.  Figures~\ref{fig:fig3} and \ref{fig:fig4} show the peak emission spectra and maps for our \oi\, FIFI-LS observations at 145 \um\, and 63 \um\, following analysis along the same path described above.

To gain further insight into the distribution of continuum and line emission in each map, we examined the continuum and line emission profiles along slices at constant Right Ascenscion and at constant Declination through the peak of continuum emission in each map.  We fit each continuum profile with a linear component to represent a distributed Galactic background plus a Gaussian to represent a localized continuum source.   For sources G302, G317, and G341, the \cii\,  line emission has a larger spatial extent than that of the continuum emission and extends beyond the edges of the maps.

Line-of-sight velocities for the sources of the \cii\,  emission can be calculated by comparison of the peak wavelength with the precisely known rest wavelength of  157.7409 $\mu$m  \citep{Cooksy1986}.  For G302, G317, and G341, the \cii\, velocities match the molecular line velocities observed in MALT90 to within $\sim$ 2 \kms.  This agreement gives us confidence in the wavelength calibration for FIFI-LS.   

\section{Discussion}
 
\subsection{Comparison with Galaxies}
\begin{figure}
\begin{center}
\includegraphics[scale=2, angle=0]{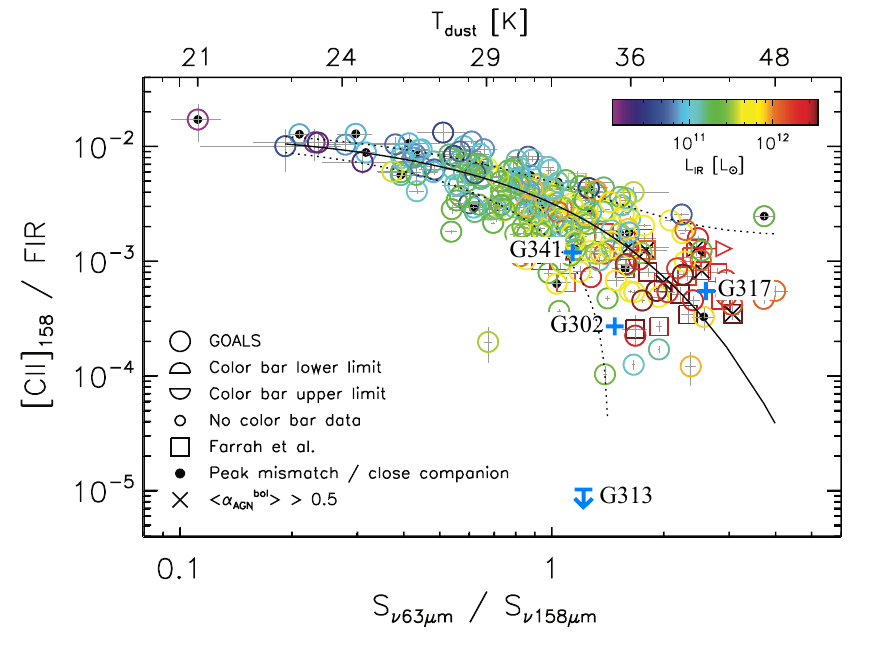} 
\caption{Figure 1, panel 4 from \citealt{Diaz-Santos2017}, showing the \cii/FIR flux ratio plotted against the ratio of the IR continuum flux density at 63 \um\, to that at 158 \um\, for the GOALS sample of galaxies, with the four sources in the current study  overplotted with blue symbols.  Three of the clumps fall within the envelope of points found for galaxies, but one source, \sourceB\, (labelled here as G313), falls at least a factor of 100 below the empirical trend. }
\label{fig:fig5}
\end{center}
\end{figure}

Here we compare the \cii\, emission from these four Galactic clumps to the galaxy-integrated \cii\, emission from 240 local luminous infrared galaxies reported by the Great Observatories All-sky LIRG Survey  \citep[GOALS,][]{Armus2009,Diaz-Santos2013,Diaz-Santos2017} .  The GOALS results are presented in Figure~\ref{fig:fig5}, which is panel 4 from Figure 1 in \cite{Diaz-Santos2017}.  To highlight the trends of extragalactic \cii\, emission with FIR luminosity and average  dust temperature, the GOALS results are presented in terms of the flux ratios  F$_{\cii}$/F$_{FIR}$ plotted versus $S_{63\mu m}/S_{158\mu m}$.  Here F$_{FIR}$ is the estimated FIR flux betwen 42.5 and 122.5 \um, following the original definition of \cite{Helou1985} and calculated as
$$F_{FIR[42.5-122.5\mu m]}=1.26 \times 10^{-14}(2.58S_{60\mu m}+S_{100\mu m})\; {\rm W ~m^{-2}}$$ 
where $S_\nu$ is the IRAS flux density  in Jy.  The flux density ratio $S_{63\mu m}/S_{158\mu m}$ is used as a proxy for dust temperature.  The T$_{dust}$ scale at the top of Figure~\ref{fig:fig5} is calculated from the $S_{63\mu m}/S_{158\mu m}$ ratios using a single-temperature modified blackbody with emissivity index 1.8.  

 The values for the four sources of the present study have been added to Figure~\ref{fig:fig5} as blue plus signs (upper-limit in the case of G313).
To calculate F$_{\cii}$/F$_{FIR}$ we searched the IRAS Point Source Catalog for matches to our source positions.  Each source had a single IRAS match to our nominal source coordinates within a 30\arcsec\, radius, at a distance ranging from 5\arcsec\, to 12\arcsec.   We used the corresponding values of $S_{60\mu m}$ and $S_{100\mu m}$ to calculate  F$_{FIR}$ for each source.  The black line in the figure is a functional fit to the data; the dotted lines indicate the one sigma dispersion from that trend line.   G313 is a distant outlier, inconsistent with the functional fit for galaxies, while the other three sources have \cii\, and FIR emission ratios consistent with the GOALS trends.

 The average dust temperatures for G302, G313, G317, and G341 found by \citealt{Guzman2015} using {\it Herschel} and ATLASGAL fluxes are 25.9, 26.9, 30.7, and 25.7 K  whereas the $S_{63\mu m}/S_{158\mu m}$ temperatures are 36.9, 36.0, 40.5, and 31.6 K.  It is not surprising that the  \cite{Guzman2015} temperatures are lower than those indicated by IRAS data alone, since their analysis 
included longer wavelength submm continuum data and would thus be more sensitive to cold dust. Moreover, the \cite{Guzman2015} analysis filtered out FIR/submm background emission on scales larger than 2.5 arcminutes.   All four of our sources were flagged as having likely contamination from infrared cirrus emission (IRAS Explanatory Supplement V.H.4, VII.H.2).  As an alternative approach, we calculated F$_{FIR}$ by scaling the total bolometric luminosity derived by \cite{Guzman2015}  and \cite{Contreras2017} to the 42.5 to 122.5 \um\,  wavelength range using a single-temperature modified blackbody with emissivity index $\beta=1.7$ and the dust temperatures derived by  \cite{Guzman2015}.   These \lfir\, flux estimates are smaller than the IRAS estimates by a factor of 1.9 to 2.4.  Using these estimates in the  $F_{\cii}$/F$_{FIR}$ determination rather than IRAS values would move our data points in Figure~\ref{fig:fig5} upward by less than twice the height of the plus symbol, which would not change our conclusions above.

The  F$_{\cii}$/F$_{FIR}$ values for our sample range from a low value of $<8.7 \times 10^{-6}$ for G313 to a high value of $1.2 \times 10^{-3}$ for G341.  For comparison, the value for the entire Orion clump is $3.8 \times 10^{-3}$ \citep{Goicoechea2015}.  The ratio varies spatially in the various physical components of Orion, ranging from  $9.3 \times 10^{-4}$ towards the Trapezium to $1.1 \times 10^{-3}$ towards the dense photodissociation region  \citep{Goicoechea2015}.

If a galaxy's \cii\, and FIR emission arises from a collection of typical Galactic clumps, these clumps should have similar \lcii/\lfir, ratios and dust temperatures as found for entire galaxies, or, at the very least, luminosity-weighted averages of the clump ensemble should be able to reproduce a galaxy's global emission.  With only four sources, it is difficult to assess whether clumps alone can account for the global \cii\, and FIR emission from galaxies.  It is of course possible that clumps like G313 are rare, but selecting such a rare clump randomly in a sample of four would be exceedingly improbable.   Perhaps clumps with higher \lcii/\lfir\, ratios are more common, and a few clumps like G313 can act to bring down the average for galaxies to the observed levels.  A more complete characterization of Galactic clumps is necessary to establish the clump properties and their distribution functions. From this small sample it is clear that Galactic dense clumps have a large variation in their \lcii/\lfir\, ratios of factors of at least 100, and this large variation suggests that dense clumps  alone are unlikely to account for a galaxy's global \cii\, and FIR emission.

\subsection{\sourceB, A Luminous Infrared Clump with No [C II] Emission}

While the \cii\, luminosity in proportion to IR luminosity for three of our four sources is in accord with the GOALS distribution for galaxies, one of our four sources, G313, lies well below the empirical relation.   The lack of \cii\, emission is surprising for such a luminous Galactic star-forming region, with a bolometric luminosity of 24,000 \Lsun.  We suggest three possibilities that might explain the apparent   lack of \cii\, emission. 

First, G313 may be in a very early ``protostellar" phase, in which the most massive young stellar objects that it hosts have not yet reached the main sequence.  In this stage of early stellar evolution, the protostar is too cold to emit ultraviolet radiation.  Thus, carbon cannot be ionized, and no \cii\, emission can be generated.  A previous 20 cm image of the region fails to detect compact 20 cm radio source coincident with G313 ($F_\nu <10$ mJy; \citealt{Roberts1999}).  However, a recent, more sensitive 1.3 cm radio continuum image of G313 with the Australia Telescope Compact Array does detect 22.18 GHz radio continuum at the 0.6 mJy level (CACHMC Survey, D. Allingham et al., in prep.)  If all of the bolometric luminosity of G313 were attributed to a single main sequence star, its mass would be $\sim 13$ to 16 \Msun\, (e.g., \citealt{Malkov2007}), and the 20 and 1.3 cm free-free flux from an optically thin, homogeneous \hii\, region surrounding such a star would be $\sim 20$ mJy (e.g., using the analysis of \citealt{JacksonKraemer1999}).  Thus, although the presence of faint radio continuum flux does imply the possible presence of a cetnral ionizing star or stars, the observed flux is inconsistent with an optically thin \hii\, region ionized by a single star with the observed bolometric luiminosity.  

Second, even if G313 contains a high-mass main sequence star, this star may be so deeply embedded in dense material that ionizing radiation cannot penetrate very far into the surrounding medium due to the high opacity of the surrounding dust to ultraviolet radiation.  Thus, if G313 is in this early evolutionary phase, the \hii\, region surrounding the star would be in a hypercompact stage, and the surrounding PDR would occupy such a small volume that no \cii\, radiation can be detected. The observed 1.3 cm radio continuum flux, although inconsistent with free-free emission from an optically thin H II region ionized by a star with the observed bolometric luminosity, is in fact consistent with an optically thick \hii\, region with an electron density $n_e > 10^4$ cm$^{-3}$, certainly plausible for a young \hii\, region in a hypercompact stage.  Furthermore, the presenece of a 6668 MHz methanol maser (141624.4-602855 \citealt{vanderWalt1995}, MMB G313.577+00.325, \citealt{Caswell2010, Green2012}) indicates an early protostellar phase of evolution (e.g., \citealt{Ellingsen2006}), but the presence of an OH maser (OH 313.5777+00.325, \citealt{Caswell1998}) implies a slighlty more advanced protostellar stage \citep{Ellingsen2006}.  Of the four clumps studied here, G313 is the only one for which methanol maser emission has been detected.  In addition, both its radio continuum and \cii\, emission are much fainter than the other three clumps. Moreover, the fact that G313 has strong 24 \um\, emission but little emission at shorter mid-IR wavelengths (and thus appears redder in Figure 2) also distinguishes it from the other three clumps and suggests a more deeply embedded, younger central source.  Finally, the fact that G313 is the only one of the four sources studied here that the MALT90 survey detected in N$_2$H$^+$  and has the largest N$_2$H$^+$/HCO$^+$ $1-0$ integrated intensity ratio of these four clumps  \citep{Rathborne2016} provides some chemical evidence for a protostellar evolutionary phase (e.g., \citealt{Hoq2013}).

Finally, it is possible that G313 is emitting \cii\, line radiation, but the line profile shows deep absorption features against the FIR continuum.  It is also possible that the flux from an intrinsically bright \cii\, line could be severely attenuated due to self-absorption by foreground \cii\, with a low excitation temperature. Because FIFI-LS has coarse spectral resolution, it is conceivable that deep enough absorption against the continuum or self-absorption against intrinsically bright line emission might cancel or diminish the integrated line flux detected within the broad FIFI-LS spectral resolution element.  Indeed, SOFIA's heterodyne GREAT instrument with $\sim$ 1 \kms\, spectral resoltuion often detects \cii\, self-absorption features toward bright star-forming regions (e.g., \citealt{Graf2015}).   \cite{Gerin2015} also detected \cii\, absorption and self-absorption features toward several star-forming regions with {\it Herschel}.  In a recent study of four molecular clouds, \cite{Guevara2020} find large optical depths in both \cii\, emission lines and in \cii\, self-absorption features, both of which have surprisingly large column densities.  \cite{Franeck2018} also predict optically thick \cii\, in their simulations of molecular cloud formation. The ubiquity of \cii\, self-absorption and absorption features indicate that optically thick, low excitation temperature \cii\, is common, and deep \cii\ absorption against a 158 \um\, continuum source is plausible. One piece of evidence in favor of this interpretation is the presence of 8 \um\, emission in two lobes to the north and south of the center of G313 (see Fig. 2, in which 8 \um\, emission is shown as green).  Since 8 \um\, emission in {\it Spitzer} images of star forming regions typically arises from a fluorescent PAH feature, these lobes may indicate ionization from a central source.  The bipolar morphology may result from outflow lobes that allow the central ionizing radiation to escape along the poles.  Indeed, our data hint at \cii\, absorption toward the northern lobe, but since the feature is faint and possibly confused with a slightly incorrect telluric correction, we cannot confirm the presence of \cii\, absorption with the current data. Future observations of G313 with higher spectral resolution could spectrally resolve the \cii\, line profile and reveal whether absorption features might diminish the integrated \cii\, flux in the FIFI-LS spectrum.

\section{Summary}
\label{sec:summary}

In an attempt to characterize the \cii\, line emission from Galactic star-forming clumps, we have observed four luminous clumps with the FIFI-LS instrument aboard SOFIA.  Three of the four clumps show bright \cii\, line emission, but the fourth, \sourceB, was not detected in \cii\, despite a rather large \lfir\, of 24,000 \Lsun.  Although it is tempting to believe that galactic-scale FIR and \cii\,  line emission might be understood as the aggregation of emission from massive-star-forming clumps, the absence of significant \cii\,  emission from \sourceB\, and the large variation of the line/continuum ratio of at least a factor of 140 challenges this simple picture.  The lack of \cii\, emission from \sourceB\, may indicate that it is in an earlier protostellar phase that lacks sufficient ultraviolet radiation to ionize carbon, or in a deeply embedded hypercompact \hii\, region phase that has an undetectably small volume of ionized carbon.  Alternatively, with FIFI's low spectral resolution of $\Delta V \sim 250$ \kms, it is also possible that \cii\, absorption against the continuum is canceling emission within the spectral resolution element.  
\newpage
\section{Acknowledgements} This research is based on observations made with the NASA/DLR Stratospheric Observatory for Infrared Astronomy (SOFIA). SOFIA is jointly operated by the Universities Space Research Association, Inc. (USRA), under NASA contract NNA17BF53C, and the Deutsches SOFIA Institut (DSI) under DLR contract 50 OK 0901 to the University of Stuttgart. Financial support for this work was provided by NASA through award 4\_0152  issued by USRA.  The Mopra radio telescope used for the MALT90 survey was part of the Australia Telescope National Facility which is funded by the Australian Government for operation as a National Facility managed by CSIRO.  The Australia Telescope Compact Array radio telescope is part of the Australia Telescope National Facility which is funded by the Australian Government for operation as a National Facility managed by CSIRO.  We thank Christof Iserlohe for providing the FLUXER software package for analysis of FIFI-LS data and Dario Fadda for assistance in calibration and telluric correction.  This research was conducted in part at the SOFIA Science Center, which is operated by the Universities Space Research Association under contract NNA17BF53C with the National Aeronautics and Space Administration.



\newpage
\movetabledown=30mm
\begin{rotatetable}
\begin{deluxetable}{lcccc}

\tabletypesize{\small}
\tablecaption{Properties of the Four Sources \label{table:properties}}
\tablehead{\colhead{Source} & \colhead{AGAL302.486-0.031}	& \colhead{AGAL313.576+0.324}	& \colhead{AGAL317.429-0.561	} & \colhead{AGAL341.702+0.051} }
\startdata
 Barycentric velocity  (\kms) &	-30.86 $\pm$0.79	&-43.97 $\pm$0.39&	29.06 $\pm$0.15	&-21.64 $\pm$0.18\\
Kinematic distance  (kpc)  &	4.6	&8.1	&15	&14.3\\
Dust temperature  T$_{dust}$(K) & 25.9 &26.9 & 30.7& 25.7\\
Mass   (10$^3$\Msun)	&0.44 $\pm$0.14	&0.79 $\pm$0.14	&3.4 $\pm$0.3	&3.9 $\pm$0.4\\
Density  ($ 10^4 cm^{-3}$) &	0.56 $\pm$0.25	&0.93 $\pm$0.23	&0.115 $\pm$0.014&0.114 $\pm$0.016\\
$L_{bol}$ ($10^3$ \Lsun)	&10.4 $\pm$3.9	&24.0 $\pm$6.4	&238.0 $\pm$53.0	&93.8 $\pm$20.4\\
FIR flux from $L_{bol}$  (W m$^{-2}$)&	1.59E-11&	1.17E-11	&3.38E-11	 &1.47E-11\\
FIR flux from IRAS  (W m$^{-2}$)&	2.40E-11&	1.67E-11	&4.76E-11	 &1.97E-11\\
L/M (\Lsun /\Msun)&23.6$\pm$ 11.6&30.4$\pm$ 9.7&70.0$\pm$ 16.8&24.05$\pm$ 5.8\\
22.18 GHz  Flux (mJy)&	15.9 $\pm$0.6&	0.62 $\pm$0.09&	44.7 $\pm$0.5  &170.9 $\pm$0.9\\
\enddata
\vskip 0.1truein

NOTES:  Velocities and kinematic distances from MALT90.  Radio continuum and H$_2$O maser data from the Complete ATCA Census of High-Mass Clumps  (Allingham et al., in prep.)

\end{deluxetable}
\end{rotatetable}

\newpage
\movetabledown40mm
\begin{rotatetable}
\begin{deluxetable}{lcccc}
\tabletypesize{\small}
\tablecaption{FIFI-LS Measurements \label{table:fifimeas}}
\tablehead{\colhead{Source} & \colhead{AGAL302.486-0.031}	& \colhead{AGAL313.576+0.324}	& \colhead{AGAL317.429-0.561	} & \colhead{AGAL341.702+0.051} }
\startdata
\sidehead{\uline{FIFI-LS 158 \um}}			
continuum peak (Jy pixel$^{-1}$)	&2.13	&3.10	&3.11&	1.25\\
total map continuum  (Jy)	&378	&396&	532	&330\\
continuum under 2D gaussian  (Jy)&256	&335	&342	&105\\
\cii\, flux peak  ($10^{-16}$ W m$^{-2}$ pixel$^{-1}$)&	0.248&	0.031	&0.656	&0.677\\
\cii\, flux total map  ($10^{-16}$ W m$^{-2}$ ) &	65.5	&3.24&	263	&237\\				
\cii\,  flux/continuum   peak (\um)&	0.096&	0.008&	0.175	&0.448\\
\cii\, flux/continuum   total  (\um)&	0.144	&0.007	&0.410	&0.597	\\			
\cii\, total flux / FIR flux	&2.7E-4	&8.7E-6&	5.5E-4	&1.2E-3\\
\sidehead{\uline{FIFI-LS 145 \um}}				
continuum peak  (Jy pixel$^{-1}$)& & &		5.3	&1.9\\
total map continuum  (Jy)		& & &		705	&443\\
continuum under 2D gaussian  (Jy)		& & &		514&	157\\
\oi\, flux peak  ($10^{-16}$ W m$^{-2}$ pixel$^{-1}$) & & &		0.242&	0.205\\
\oi\,flux total map  ($10^{-16}$ W m$^{-2}$ )  & & &		38.3	&30.0\\				
\oi\,flux/continuum   peak (\um)		& & &		0.032	&0.078\\
\oi\,flux/continuum   total  (\um)		& & &		0.038&	0.048\\				
\oi\, total flux / FIR flux 		& & &		0.00011&	0.00020\\
\sidehead{\uline{FIFI-LS 63 \um}}				
continuum peak (Jy pixel$^{-1}$)&	2.6	&3.8	&7.8&	1.0\\
total map continuum  (Jy)	&680.0	&648.0	&1354.0&	334.0\\
continuum under 2D gaussian  (Jy)&	379.0	&407.0&	890.0&	120.0\\
\enddata

\end{deluxetable}
\end{rotatetable}

\newpage
\clearpage

\newpage
\begin{thebibliography}{dummy}
\bibitem[Allamandola et al.(1989)]{Allamandola1989}Allamandola, L.~J., Tielens, A.~G.~G.~M., \& Barker, J.~R.\ 1989, \apjs, 71, 733
\bibitem[Armus et al.(2009)]{Armus2009} Armus, L., Mazzarella, J.~M., Evans, A.~S., et al.\ 2009, \pasp, 121, 559
\bibitem[Beuther et al.(2014)]{Beuther2014} Beuther, H., Ragan, S.~E., Ossenkopf, V., et al.\ 2014, \aap, 571, A53
\bibitem[Capak et al.(2015)]{Capak2015} Capak, P.~L., Carilli, C., Jones, G., et al.\ 2015, \nat, 522, 455
\bibitem[Carey et al.(1998)]{Carey1998} Carey, S.~J., Clark, F.~O., Egan, M.~P., et al.\ 1998, \apj, 508, 721
\bibitem[Caswell(1998)]{Caswell1998} Caswell, J.~L.\ 1998, \mnras, 297, 215
\bibitem[Caswell et al.(2010)]{Caswell2010} Caswell, J.~L., Fuller, G.~A., Green, J.~A., et al.\ 2010, \mnras, 404, 1029
\bibitem[Contreras et al.(2013)]{Contreras2013} Contreras, Y., Schuller, F., Urquhart, J.~S., et al.\ 2013, \aap, 549, A45
\bibitem[Contreras et al.(2017)]{Contreras2017} Contreras, Y., Rathborne, J.~M., Guzm{\'a}n, {\'A}., et al.\ 2017, \mnras, 466, 340
\bibitem[Cooksy et al.(1986)]{Cooksy1986} Cooksy, A.~L., Blake, G.~A., \& Saykally, R.~J.\ 1986, \apjl, 305, L89
\bibitem[D{\'{\i}}az-Santos et al.(2013)]{Diaz-Santos2013} D{\'{\i}}az-Santos, T., Armus, L., Charmandaris, V., et al.\ 2013, \apj, 774, 68
\bibitem[D{\'{\i}}az-Santos et al.(2017)]{Diaz-Santos2017} D{\'{\i}}az-Santos, T., Armus, L., Charmandaris, V., et al.\ 2017, \apj, 846, 32
\bibitem[Ellingsen(2006)]{Ellingsen2006} Ellingsen, S.~P.\ 2006, \apj, 638, 241
\bibitem[Fischer et al.(2018)]{Fischer2018} Fischer, C., Beckmann, S., Bryant, A., et al.\ 2018, Journal of Astronomical Instrumentation , 7, 1840003-556 
\bibitem[Foster et al.(2011)]{Foster2011} Foster, J.~B., Jackson, J.~M., Barnes, P.~J., et al.\ 2011, \apjs, 197, 25
\bibitem[Foster et al.(2013)]{Foster2013} Foster, J.~B., Rathborne, J.~M., Sanhueza, P., et al.\ 2013, \pasa, 30, e038
\bibitem[Gao \& Solomon(2004)]{Gao2004} Gao, Y., \& Solomon, P.~M.\ 2004, \apj, 606, 271
\bibitem[Franeck et al.(2018)]{Franeck2018} Franeck, A., Walch, S., Seifried, D., et al.\ 2018, \mnras, 481, 4277
\bibitem[Gerin et al.(2015)]{Gerin2015} Gerin, M., Ruaud, M., Goicoechea, J.~R., et al.\ 2015, \aap, 573, A30
\bibitem[Goicoechea et al.(2015)]{Goicoechea2015} Goicoechea, J.~R., Teyssier, D., Etxaluze, M., et al.\ 2015, \apj, 812, 75
\bibitem[Graf et al.(2015)]{Graf2015} Graf, U.~U., Simon, R., Stutzki, J., et al.\ 2015, EAS Publications Series, 189
\bibitem[Green et al.(2012)]{Green2012} Green, J.~A., Caswell, J.~L., Fuller, G.~A., et al.\ 2012, \mnras, 420, 3108
\bibitem[Guevara et al.(2020)]{Guevara2020} Guevara, C., Stutzki, J., Ossenkopf-Okada, V., et al.\ 2020, \aap, 636, A16
\bibitem[Gullberg et al.(2015)]{Gullberg2015} Gullberg, B., De Breuck, C., Vieira, J.~D., et al.\ 2015, \mnras, 449, 2883
\bibitem[Guzm{\'a}n et al.(2015)]{Guzman2015} Guzm{\'a}n, A.~E., Sanhueza, P., Contreras, Y., et al.\ 2015, \apj, 815, 130	
\bibitem[Graf et al.(2015)]{Graf2015} Graf, U.~U., Simon, R., Stutzki, J., \& G{\"u}sten, R.\ 2015, EAS Publications Series, 75, 189 
\bibitem[Helfer \& Blitz(1997)]{HelferBlitz1997} Helfer, T.~T., \& Blitz, L.\ 1997, \apj, 478, 233
\bibitem[Helou et al.(1985)]{Helou1985} Helou, G., Soifer, B.~T., \& Rowan-Robinson, M.\ 1985, \apjl, 298, L7
\bibitem[Hennebelle et al.(2001)]{Hennebelle2001} Hennebelle, P., P{\'e}rault, M., Teyssier, D., et al.\ 2001, \aap, 365, 598
\bibitem[Hoq et al.(2013)]{Hoq2013} Hoq, S., Jackson, J.~M., Foster, J.~B., et al.\ 2013, \apj, 777, 157
\bibitem[Ibar et al.(2015)]{Ibar2015} Ibar, E., Lara-L{\'o}pez, M.~A., Herrera-Camus, R., et al.\ 2015, \mnras, 449, 2498
\bibitem[Jackson \& Kraemer (1999)]{JacksonKraemer1999} Jackson, J.~M. \& Kraemer, K.~E.\ 1999, \apj, 512, 260
\bibitem[Jackson et al.(2013)]{Jackson2013}Jackson, J.~M., Rathborne, J.~M., Foster, J.~B., et al.\ 2013, \pasa, 30, e057
\bibitem[Kauffmann et al.(2017)]{Kauffmann2017} Kauffmann, J., Goldsmith, P.~F., Melnick, G., et al.\ 2017, \aap, 605, L5
\bibitem[Kimball et al.(2015)]{Kimball2015} Kimball, A.~E., Lacy, M., Lonsdale, C.~J., \& Macquart, J.-P.\ 2015, \mnras, 452, 88
\bibitem[Luhman et al.(1998)]{Luhman1998} Luhman, M.~L., Satyapal, S., Fischer, J., et al.\ 1998, \apjl, 504, L11
\bibitem[Madden(2000)]{Madden2000} Madden, S.~C.\ 2000, \nar, 44, 249
\bibitem[Malkov(2007)]{Malkov2007} Malkov, O.~Y.\ 2007, \mnras, 382, 1073
\bibitem[McQuinn et al.(2002)]{McQuinn2002} McQuinn, K.~B.~W., Simon, R., Law, C.~J., et al.\ 2002, \apj, 576, 274
\bibitem[Molinari et al.(2010)]{Molinari2010} Molinari, S., Swinyard, B., Bally, J., et al.\ 2010, \aap, 518, L100
\bibitem[Pabst et al.(2019)]{Pabst2019} Pabst, C., Higgins, R., Goicoechea, J.~R., et al.\ 2019, \nat, 565, 618
\bibitem[Pety et al.(2017)]{Pety2017} Pety, J., Guzm{\'a}n, V.~V., Orkisz, J.~H., et al.\ 2017, \aap, 599, A98
\bibitem[Pineda et al.(2014)]{Pineda2014} Pineda, J.~L., Langer, W.~D., \& Goldsmith, P.~F.\ 2014, \aap, 570, A121
\bibitem[Rathborne et al.(2006)]{Rathborne2006} Rathborne, J.~M., Jackson, J.~M., \& Simon, R.\ 2006, \apj, 641, 389
\bibitem[Rathborne et al.(2016)]{Rathborne2016}Rathborne, J.~M., Whitaker, J.~S., Jackson, J.~M., et al.\ 2016, \pasa, 33, e030
\bibitem[Roberts et al. (1999)]{Roberts1999} Roberts, M~S.~E., Romani, R.~W., Johnston, S., \& Green, A.~J. \ 1999, \apj, 515, 712
\bibitem[Sargsyan et al.(2012)]{Sargsyan2012} Sargsyan, L., Lebouteiller, V., Weedman, D., et al.\ 2012, \apj, 755, 171 
\bibitem[Schuller et al.(2009)]{Schuller2009} Schuller, F., Menten, K.~M., Contreras, Y., et al.\ 2009, \aap, 504, 415
\bibitem[Stacey et al.(1993)]{Stacey1993} Stacey, G.~J., Jaffe, D.~T., Geis, N., et al.\ 1993, \apj, 404, 219
\bibitem[Stacey et al.(2010)]{Stacey2010} Stacey, G.~J., Hailey-Dunsheath, S., Ferkinhoff, C., et al.\ 2010, \apj, 724, 957
\bibitem[Stephens et al.(2016)]{Stephens2016} Stephens, I.~W., Jackson, J.~M., Whitaker, J.~S., et al.\ 2016, \apj, 824, 29
\bibitem[Temi et al.(2018)]{Temi2018} Temi, P., Hoffman, D., Ennico, K., et al.\ 2018, Journal of Astronomical Instrumentation, 7, 1840011-186
\bibitem[van der Walt et al.(1995)]{vanderWalt1995} van der Walt, D.~J., Gaylard, M.~J., \& MacLeod, G.~C.\ 1995, \aaps, 110, 81
\bibitem[Whitaker et al.(2017)]{Whitaker2017} Whitaker, J.~S., Jackson, J.~M., Rathborne, J.~M., et al.\ 2017, \aj, 154, 140
\bibitem[Wu et al.(2005)]{Wu2005} Wu, J., Evans, N.~J., II, Gao, Y., et al.\ 2005, \apjl, 635, L173
\bibitem[Wu et al.(2010)]{Wu2010} Wu, J., Evans, N.~J., Shirley, Y.~L., et al.\ 2010, \apjs, 188, 313
\bibitem[Young et al.(2012)]{Young2012} Young, E.~T., Becklin, E.~E., Marcum, P.~M., et al.\ 2012, \apjl, 749, L17




 
\end{thebibliography}
\end{document}